\documentclass[12pt,onecolumn]{emulateapj}
\pdfoutput=1
\usepackage{amssymb}
\usepackage{amsmath}
\usepackage{epsfig}
\usepackage{subfigure}
\usepackage{mathrsfs}
\usepackage{longtable}
\usepackage[usenames,dvipsnames]{xcolor}
\usepackage{mathtools}
\usepackage{relsize}
\usepackage{hyperref}
\begin{document}
 \renewcommand{\thefigure}{\arabic{figure}}
\newcommand{\noj}{}

\newcommand{\be}{\begin{equation}}
\newcommand{\ee}{\end{equation}}
\newcommand{\bea}{\begin{eqnarray}}
\newcommand{\eea}{\end{eqnarray}}
\newcommand{\lsim}{\mathrel{\hbox{\rlap{\lower.55ex\hbox{$\sim$}} \kern-.3em \raise.4ex \hbox{$<$}}}}
\newcommand{\gsim}{\mathrel{\hbox{\rlap{\lower.55ex\hbox{$\sim$}} \kern-.3em \raise.4ex \hbox{$>$}}}}
\newcommand{\grad}{\ensuremath{\vec{\nabla}}}
\newcommand{\adotoa}{\ensuremath{{\cal H}}} 
\newcommand{\Uc}{\ensuremath{{\cal U}}}
\newcommand{\Vc}{\ensuremath{{\cal V}}}
\newcommand{\Jc}{\ensuremath{{\cal J}}}
\newcommand{\Mc}{\ensuremath{{\cal M}}}

\newcommand{\unit}[1]{\ensuremath{\, \mathrm{#1}}}
%%%%%%%%%%%%% colortext comments %%%%%%%%%%%%%%
\newcommand{\tkRH}[1]{\textcolor{blue}{#1}}		  % RH
%%%%%%%%%%%%%%%%%%%%%%%%%%%%%%%%%%%%%%%%%%
\renewcommand{\arraystretch}{1.35}
\baselineskip=11.5pt

\title{Planck Data Reconsidered}

\author{David N. Spergel$^1$, Raphael Flauger$^{2,3}$ \& Ren\'{e}e Hlo\v{z}ek$^1$}
     \affiliation{$^1$Department of Astrophysical Sciences, Princeton University, Princeton, NJ 08544, USA \\
     $^2$Institute for Advanced Study, Einstein Drive, Princeton, NJ 08540, USA\\
     $^3$CCPP, New York University, New York, NY 10003, USA}

\begin{abstract}
\baselineskip=11.5pt
The tension between the best fit parameters derived by the Planck team and a number of other astronomical measurements suggests either systematics in the astronomical measurements, systematics in the Planck data, the need for new physics, or a combination thereof.
We re-analyze the Planck data and find that the $217\,\text{GHz}\times 217\,\text{GHz}$ detector set spectrum used in the Planck analysis is responsible for some of this tension. We use a map-based foreground cleaning procedure, relying on a combination of 353 GHz and 545 GHz maps to reduce residual foregrounds in the intermediate frequency maps used for cosmological inference. For our baseline data analysis, which uses 47\% of the sky and makes use of both 353 and 545 GHz data for foreground cleaning, we find the $\Lambda$CDM cosmological parameters $\Omega_c h^2 = 0.1170 \pm 0.0025$, \mbox{$n_s = 0.9686 \pm 0.0069$}, $H_0 = 68.0 \pm 1.1\,\mathrm{km}~\mathrm{s}^{-1}\mathrm{Mpc}^{-1}$, $\Omega_b h^2 = 0.02197 \pm 0.00026$, $\ln 10^{10}A_s = 3.082 \pm 0.025$, and $\tau = 0.090 \pm 0.013 $.
While in broad agreement with the results reported by the Planck team, these revised parameters imply a universe with a lower matter density of $\Omega_m=0.302\pm0.015$, and parameter values generally more consistent with pre-Planck CMB analyses and astronomical observations. We compare our cleaning procedure with the foreground modeling used by the Planck team and find good agreement. The difference in parameters between our analysis and that of the Planck team is mostly due to our use of cross-spectra from the publicly available survey maps instead of their use of the detector set cross-spectra which include pixels only observed in one of the surveys. We show evidence suggesting residual systematics in the detector set spectra used in the Planck likelihood code, which is substantially reduced for our spectra. Using our cleaned survey cross-spectra, we recompute the limit on neutrino species and find $N_\text{eff} = 3.34 \pm 0.35$.   We also recompute limits on the $n_s-r$ plane, and neutrino mass constraints.
\end{abstract}
     \maketitle
\global\csname @topnum\endcsname 0
\global\csname @botnum\endcsname 0

\section{Introduction}

Observations of the cosmic microwave background are our most powerful probe of the physical conditions in the early universe and can provide precise and accurate determinations of cosmological parameters.  Over the past decade, analyses of the data from the Wilkinson Microwave Anisotropy Probe \citep{spergel2003, spergel2006, dunkley2009, komatsu2011, hinshaw2012} have established a now standard six parameter $\Lambda$CDM model for cosmology.  In combination with ever-improving ground-based cosmic microwave background data, the most recent being the Atacama Cosmology Telescope (ACT) \citep{sievers/etal:2013} and the South Pole Telescope (SPT) \citep{story2012}, these measurements have yielded increasingly accurate determinations of cosmic parameters.

The eagerly anticipated Planck satellite was the next major step forward in the study of the cosmic microwave background \citep{planck:overview}. The data are of exceptional quality, with excellent noise properties and above-expected performance of the satellite and its components. The data at 30, 44, 70, 100, 143, 217, 353, 545 and 857 GHz provide maps which are not only useful for cosmology, but, through their increased angular resolution, have also improved our understanding of the emission from astrophysical sources in the universe.

One of the most surprising results from the Planck satellite is the measurement of cosmological parameters based on its three central frequencies, 100 GHz, 143 GHz and 217 GHz \citep{planck:likelihood,planck:parameters} 
that finds a higher matter density, $\Omega_m =0.315^{+0.016}_{-0.018}$,
a higher amplitude of matter fluctuations, $\sigma_8 (\Omega_m/0.27)^{0.3} = 0.87 \pm 0.02$,  
and a lower Hubble constant, $H_0 = 67.3 \pm 1.2~\mathrm{km}~\mathrm{s}^{-1}\mathrm{Mpc}^{-1}$ than previous analyses of CMB data \citep{calabrese/etal:2013}.
These new Planck values are all around $2$-$3\,\sigma$ discrepant with various astronomical measurements including measurements of the Hubble constant using distance ladders \citep{riess2011,freedman2012}, measurements of the matter density using
supernova \citep{conley2011} and measurements of the amplitude of matter fluctuations using a variety of techniques: gravitational lensing \citep{huff/etal:2011,heymans2013}, two and three-point statistics of the SZ map \citep{planck:SZ,crawford/etal:2013,wilson/etal:2012} and cluster counts with mass determinations based on either SZ measurements \citep{reichardt/etal:2012,planck:clusters}, X-ray measurements \citep{vikhlinin/etal:2009,hajian/etal:2013}, or optical clustering and lensing \citep{cacciato/etal:2013}. These deviations suggest either systematics in the astronomical probes, a failure of the ``simple" six parameter $\Lambda$CDM model, or systematics in the analysis of the Planck data.

The conclusions of the Planck paper rest at least in part on their analysis of the 217 GHz maps. This is highlighted by the sensitivity of the best-fit $n_s$ to the inclusion of the $217 \times 217$ power spectrum shown in Appendix C of Paper XVI \citep{planck:parameters}. At first sight, the observed shift could be due to the additional data provided by the $217 \times 217$ power spectrum. After all, the 217 GHz channel has the highest resolution among the channels used for the Planck analysis. However, the power spectrum made from the 217 GHz data shows a strong departure from the standard model at $\ell \simeq1800$ which is not seen at other frequencies, and is thus not of cosmological origin. In addition, the 217~GHz spectra fail a number of null tests shown by the Planck team in Figures 29 and 30 of Paper VI \citep{planck:hfi}. This suggests that the dependence on the $217 \times 217$ data may in part be due to systematics. In Section~\ref{sec:consistency}, we show that there is not only a significant difference between the Planck $217\times217$ spectrum and the $100\times100$, $143\times143$, and $143\times217$ spectra around $\ell\simeq 1800$ but also for $\ell\gtrsim 2000$, and that the removal of the $217\times217$ spectrum from the Planck likelihood analysis alters best-fit parameters by more than expected from simulations.\footnote{From the revised {\em Planck} publication~\citep{planck:parameters} we know that the difference around $\ell\simeq 1800$ arises from the incomplete removal of the 4K cooler line. In addition, the revised publication~\citep{planck:parameters} acknowledges an error in the ordering of 217 GHz beam transfer functions that leads to changes in the 217 GHz spectra by a few $\mu K^2$ for $\ell>2000$. To what extent this is related to the excess noted here is difficult to assess. The corrected spectra are not publicly available, and this may be compensated in the likelihood analysis by a shift in the parameters of the foreground model.}

In Section~\ref{sec:clean}, we make use of the higher frequency Planck data at 353 GHz and 545 GHz to recompute the CMB power spectrum and perform an analysis that is much less sensitive to the foreground model. 
At small angular scales, dusty galaxies are a significant foreground and contribute more to the variance
of these maps than the cosmic microwave background fluctuations. Planck's sensitive high-frequency channels trace the sub-millimeter fluctuations in the dusty galaxy distribution well.  By minimizing the variance in linear combinations of the lower frequency maps (100, 143 and 217 GHz) with a combination of the 353 or 545 GHz maps: we ``clean'' the maps.  We also use the multi-frequency data to define masks that are both unbiased (relative to the CMB) and use a larger fraction of the sky than the masks used by the Planck team. If we restrict ourselves to the same masks as used by the Planck team, the analysis based on our cleaned spectra agrees well with our analysis of uncleaned survey cross-spectra with the same foreground model as used by the Planck team. However, our cleaning procedure allows us to use significantly more of the sky without biasing the spectra or parameters.   
%Because these foregrounds are very large and have a non-trivial dependance on multipoles and the Planck errors on fluctuation amplitude are very small, the foreground fluctuations must be modeled
%to high precision in order to not bias parameter determination. 

In Section~\ref{sec:parameters}, we use the new maps to recompute cosmological parameters.  
%We begin by showing that for the standard Planck analysis, the value of $H_0$ and $\Omega_m$ is also sensitive to the inclusion of the $217 \times 217$ GHz data.
For the $\Lambda$CDM model, the parameters systematically shift towards the results of earlier analyses based on WMAP, ACT and SPT \citep{calabrese/etal:2013}, decreasing the tension with other astronomical measurements. In this section, we also consider the implications of the new parameters for neutrino properties and for inflationary models.

In Section~\ref{sec:compare}, we discuss what might be the origin of the shifts. We show that our cosmological parameters are neither sensitive to the sky fraction used nor to whether we compute the power spectrum from the cleaned maps or instead implement a power spectrum-based correction as is done in the Planck team analysis. We compare our cross-spectra to the detector set spectra used in the Planck analysis. The difference seems to be primarily due to an offset between the released Planck spectra (which we will refer to as `CAMspec' spectra) and the cross-spectra, suggesting that difference spectra computed between the Planck detector set maps and the survey maps will fail null tests. Some of the shifts when removing the $217\times217$ data remain even when using the survey cross-spectra. However, the shifts are smaller if the data are cleaned using our procedure and consistent with expectations based on simulations.

\section{The 217$\times$ 217 Power Spectrum: A Fly in the Ointment?}
\label{sec:consistency}
\begin{figure}[t]
\begin{center}
\includegraphics[width=0.9\textwidth]{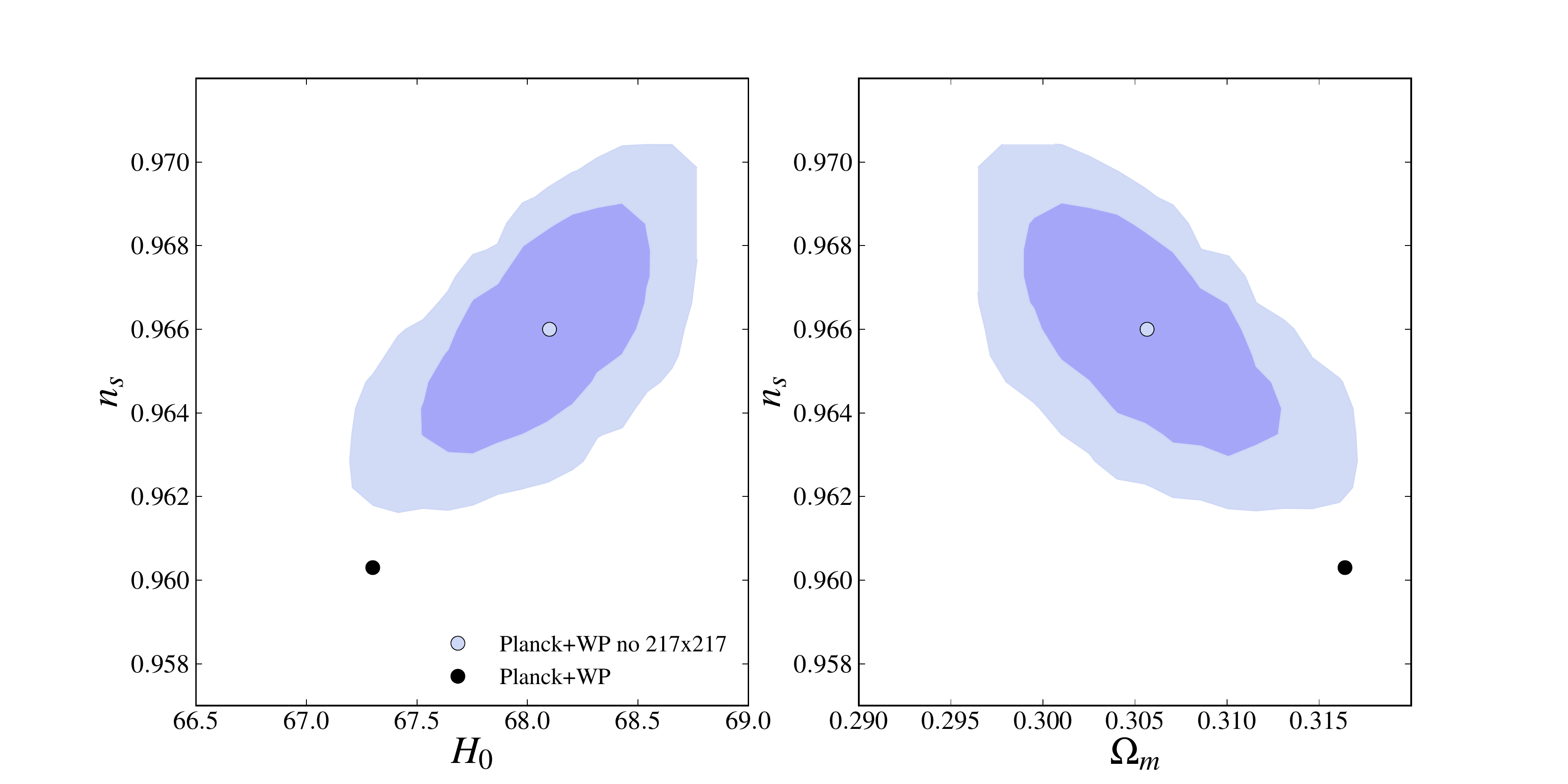}
    \caption{Means for $n_s$, $\Omega_M$ and $H_0$ derived from the publicly available Planck likelihood code (black) and without the $217\times217$ data (gray), both for the Planck+WP data set. The contours indicate the regions that contain $68$ and $95\%$ of our $500$ simulations with $217\times217$ spectra drawn from the conditional probability~\eqref{eq:predict}.
    \label{fig:sim217}}
\end{center}
\end{figure}
The $217 \times 217$ detector set spectra are responsible for a significant amount of the shift in cosmological parameters reported by Planck. To illustrate this, we have used the publicly released Planck likelihood code to run Monte-Carlo Markov Chains for the Planck data (Planck+WP) as used by the Planck team and without the $217\times 217$ data.  Table~\ref{tab:wmap} compares the results of these analyses to the parameters derived from an analysis based on WMAP9 and ACT data \citep{calabrese/etal:2013}.
\begin{table} [h]
\begin{center}
\caption{Planck versus pre-Planck $\Lambda$CDM Cosmological Parameters\label{tab:wmap}}
\begin{tabular}{|c||c|c|c|}
\hline
& Planck Analysis & No $217 \times 217$ & WMAP9+ACT\\
\hline
10 $\Omega_c h^2$ & 1.199$\pm$0.026 & 1.181$\pm$ 0.027 & 1.146$\pm$0.043\\
$n_s$ &0.9603$\pm$0.0073 & 0.9661$\pm$0.0077 & 0.973 $\pm$ 0.011 \\
$H_0$ &67.3$\pm$1.2 & 68.1$\pm$1.2 & 69.7 $\pm$ 2.0\\
 100 $\Omega_b h^2$ &2.205$\pm$ 0.028& 2.226 $\pm$0.029 & 2.260 $\pm $0.041\\
  $\Omega_m$ &0.315$\pm$ 0.016& 0.305 $\pm$0.016 & 0.284 $\pm $0.024\\
 \hline
\end{tabular}
\end{center}
\end{table}%
Without the $217\times217$ data, the Planck results have 20-36\% smaller errors than the WMAP9+ACT numbers, and the values for $\Omega_c h^2$, $H_0$ and $n_s$ agree within $1\,\sigma$.  The inclusion of the $217\times217$ power spectrum further decreases the errors by $\sim 15\%$, but shifts the resulting mean values for $H_0, n_s$ and  $\Omega_c h^2$ by about $1\,\sigma$.\footnote{In comparing the numbers in Table~\ref{tab:wmap} one should keep in mind that the Planck analysis assumed the sum of neutrino masses to be $0.06\,eV$, while this was set to zero in previous analyses (such as~\citep{calabrese/etal:2013} ). As shown in~\citep{planck:parameters}, this leads to a small shift in parameters. To facilitate comparison with the Planck results, we set the sum of the neutrino masses to $0.06\,eV$ as well.} 
Motivated by these shifts, we have explored whether the reported  $217 \times 217$ power spectra are consistent with the other spectra. To quantify magnitude of shifts that are expected when adding the $217 \times 217$ data, we have performed simulations. We use the inverse covariance matrix in the CAMspec likelihood code, $\mathcal{D}$, to derive the conditional probability for $217 \times 217$ power spectra given the measured $100\times100$, $143\times143$, and $143\times217$ power spectra
\begin{eqnarray}
p(Y_\alpha|Y_j) &=& \sqrt{{\rm det}\left(\frac{\mathcal{D}_{\alpha \beta}}{2\pi}\right)} \times \exp \left[ -\frac{1}{2} (Y_\alpha - \langle Y_\alpha\rangle)\mathcal{D}_{\alpha \beta}(Y_\beta - \langle Y_\beta\rangle)\right]\qquad\text{with}\qquad\langle Y_\alpha\rangle = -\mathcal{D}_{\alpha \beta}^{-1} \mathcal{D}_{\beta_j} Y_j\,. \label{eq:predict}
\end{eqnarray}
Here $Y_\alpha$ is the difference between the $217\times 217$ spectrum and the fiducial model, $Y_j$ are the difference between the $100 \times 100$, $143 \times 143$, and $143 \times 217$ spectra and their fiducial models, and $\mathcal{D}_{\alpha \beta}^{-1}$ denotes the inverse of the $217\times217\times217\times217$ block of the inverse covariance matrix.
\begin{figure}[h]
\begin{center}
$\begin{array}{@{\hspace{-0.0in}}l}
\includegraphics[width=4.5in, trim=0mm 0mm 0mm 0mm, clip]{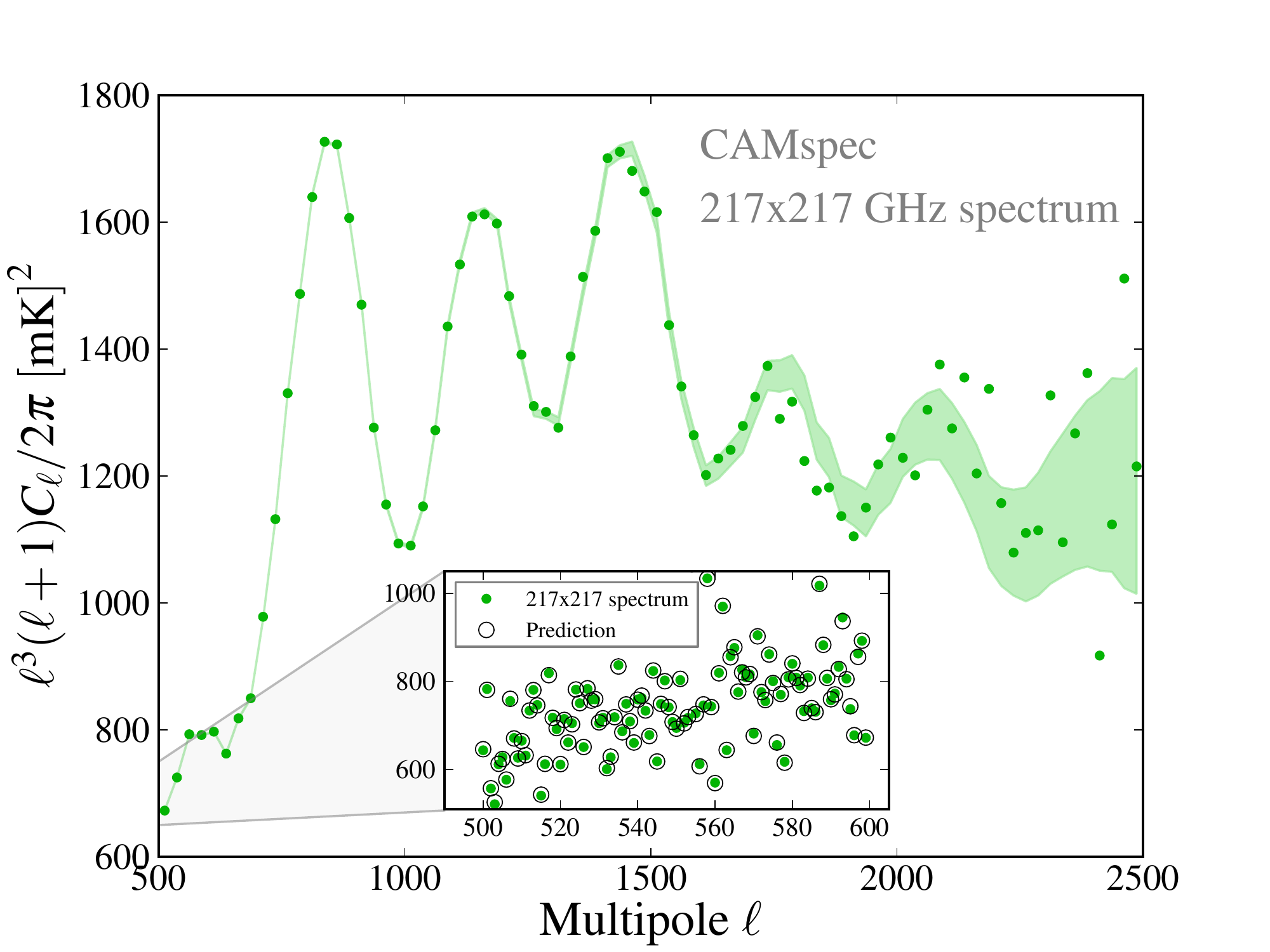}\\ [0.0cm]
\end{array}$
\vspace{-0.1in}
    \caption{Predicted versus observed power spectra: This figure shows the difference between the published $217\times217$ spectrum binned with $\Delta \ell = 25$ (points) and the spectrum predicted from the published Planck $100\times 100$, $143\times 143$, and $143\times217$ spectra (filled band indicates 68\% CL). Note that the points are systematically below the model for $1700<\ell <1900$ and above the model for $2100<\ell< 2400.$ In the boxed insert, we show the predicted spectrum and observed spectrum for each multipole between $500 < \ell< 600$ where predicted and measured spectra agree remarkably well.
    \label{fig:consistency}}
\vspace{0.1in}
  \end{center}
  \end{figure}
We then draw $217\times217$ spectra from this distribution, combine them with the measured $100 \times 100$, $143 \times 143$, $143 \times 217$ spectra and derive the cosmological parameters for each simulation. The result for $500$ simulations is shown in Figure~\ref{fig:sim217}. Uncertainties in the beams are not included in the simulations, and the details depend somewhat on whether means or best-fits are considered, but it is safe to say that the shifts are somewhat larger than expected. The different frequencies agree very well when the multipole range is restricted to $\ell<1500$ and an analysis restricted to this range leads to values of cosmological parameters consistent with WMAP9+ACT. In \citep{planck:inflation}, the Planck team identified a feature near $\ell\simeq 1800$ with a significance of $2.4-3.1 \sigma$. Replacing the data in this range by simulations, this feature does not, however, lead to significant shifts in cosmological parameters. This is partly because it is only present in the $217\times 217$ data, and partly because it is hard to accommodate such a feature by adjusting parameters in $\Lambda$CDM.  As was already noted by the Planck team (see e.g., Figure B.3 in) \citep{planck:parameters}, a comparable shift is observed when removing data for $\ell> 2000$ from the fit. It is tempting to conclude that the excess is responsible for the observed shift, but it should be kept in mind that removing the $\ell > 2000$ data from the fit changes the weighting so that other ranges of multipoles may also be important to understand the origin of these shifts. 
 
Another way to test the consistency between the different spectra is to show the prediction for the $217\times217$ spectra together with the measured spectra. This is significantly easier because both the mean and the variance are known analytically from~\eqref{eq:predict}. 
Figure \ref{fig:consistency} shows this comparison of the predicted $217 \times 217$ spectra with the observed spectra.  There are two noticeable deviations: the points are systematically low at $\ell=1800$ as discussed,  and systematically high at $\ell > 2000$, a multipole range where a number of the 217 GHz detectors fail their null tests \citep{planck:hfi}. 
%The $\ell=1800$ feature is suggestively near one of the resonances of the 4K cooler line \citep{planck:hfi}; however, this cooler line should have been removed in the HFI processing.
The $\ell=1800$ feature is suggestively near one of the resonances of the 4K cooler line \citep{planck:hfi}; however, this cooler line should have been removed in the HFI processing.\footnote{At the time this work was submitted, the origin of the feature had not been established. In the revised {\em Planck} publication~\citep{planck:parameters} it was identified as a systematic associated with the incomplete removal of the time-variable electromagnetic interference between the 4K cooler electronics and the read-out electronics from the time-ordered data. Furthermore, the revised publication mentions an error in the ordering of beam transfer functions causing a change in the 217 GHz spectrum by a few ($\mu K^2$)~\citep{planck:parameters}, just the right amount to account for the systematic difference in the spectra at $\ell > 2000$. }

\section{Planck Spectrum Revisited}
\label{sec:clean}

Motivated by the discrepancies discussed in Section \ref{sec:consistency}, we have performed an analysis based on the publicly available Planck data. Since the Planck team has only released survey and halfring maps rather than the detector set maps used in their power spectrum analysis, we cannot directly reproduce their analysis (see Section~\ref{sec:observing} for more discussion on the halfring maps). On the one hand this means we are using a subset of the Planck data in the power spectrum measurements, and the error bars (for the same $f_\text{sky}$) are larger in our analysis.  On the other hand, our approach of using only survey cross-spectra is much more robust to systematics that are common between detectors observing the sky at the same time. In fact, in the analysis of ground-based CMB experiments, most analyses use only cross-spectra of maps of the sky observed at different times, e.g. \citep{das/etal:2013} as these are less prone to common-mode systematics.

\begin{figure} [ht]
\label{fig:mask}
  \begin{center}
  \hskip -0.0in
    \includegraphics[width=3.35in]{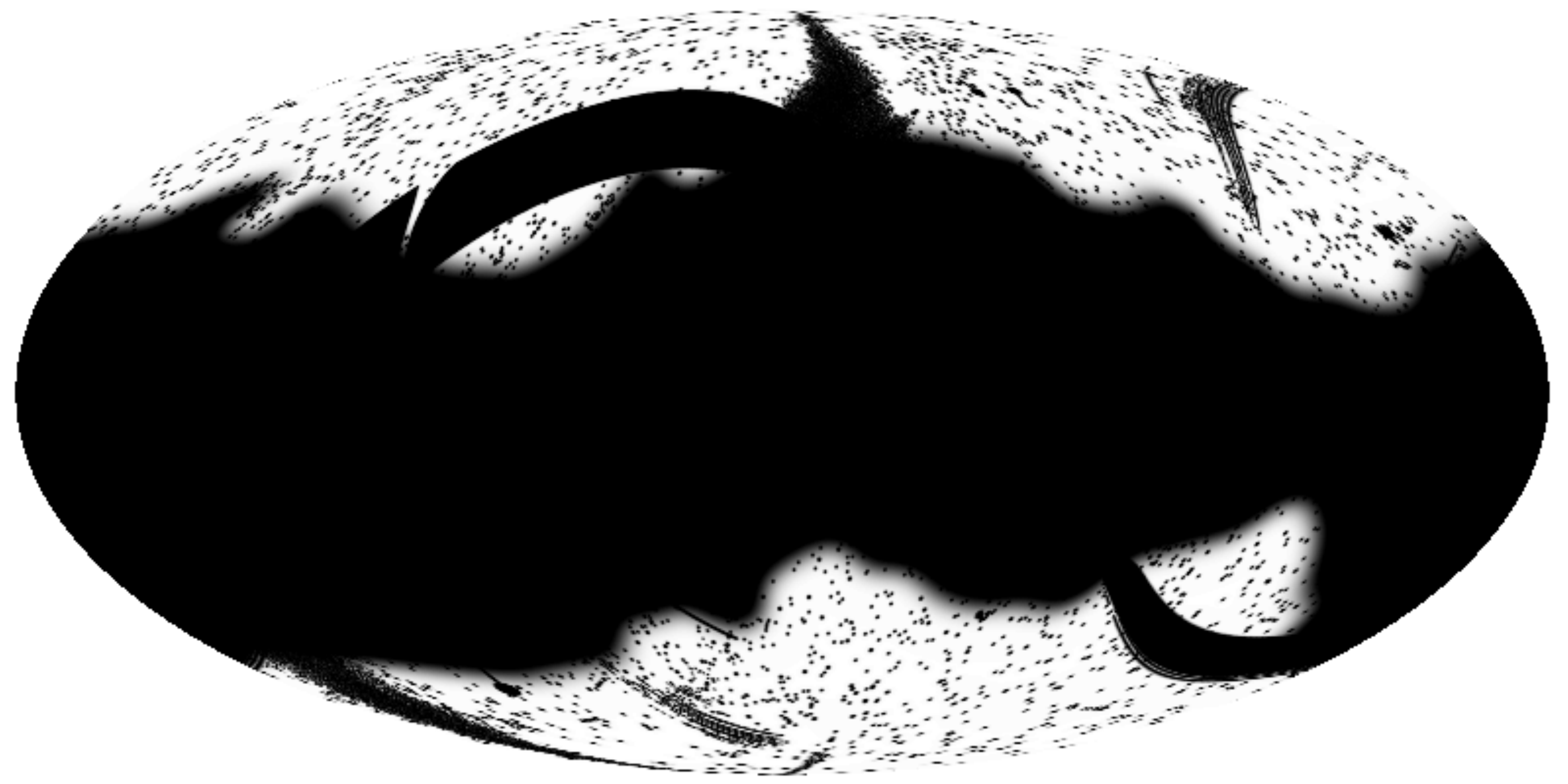}
        \includegraphics[width=3.35in]{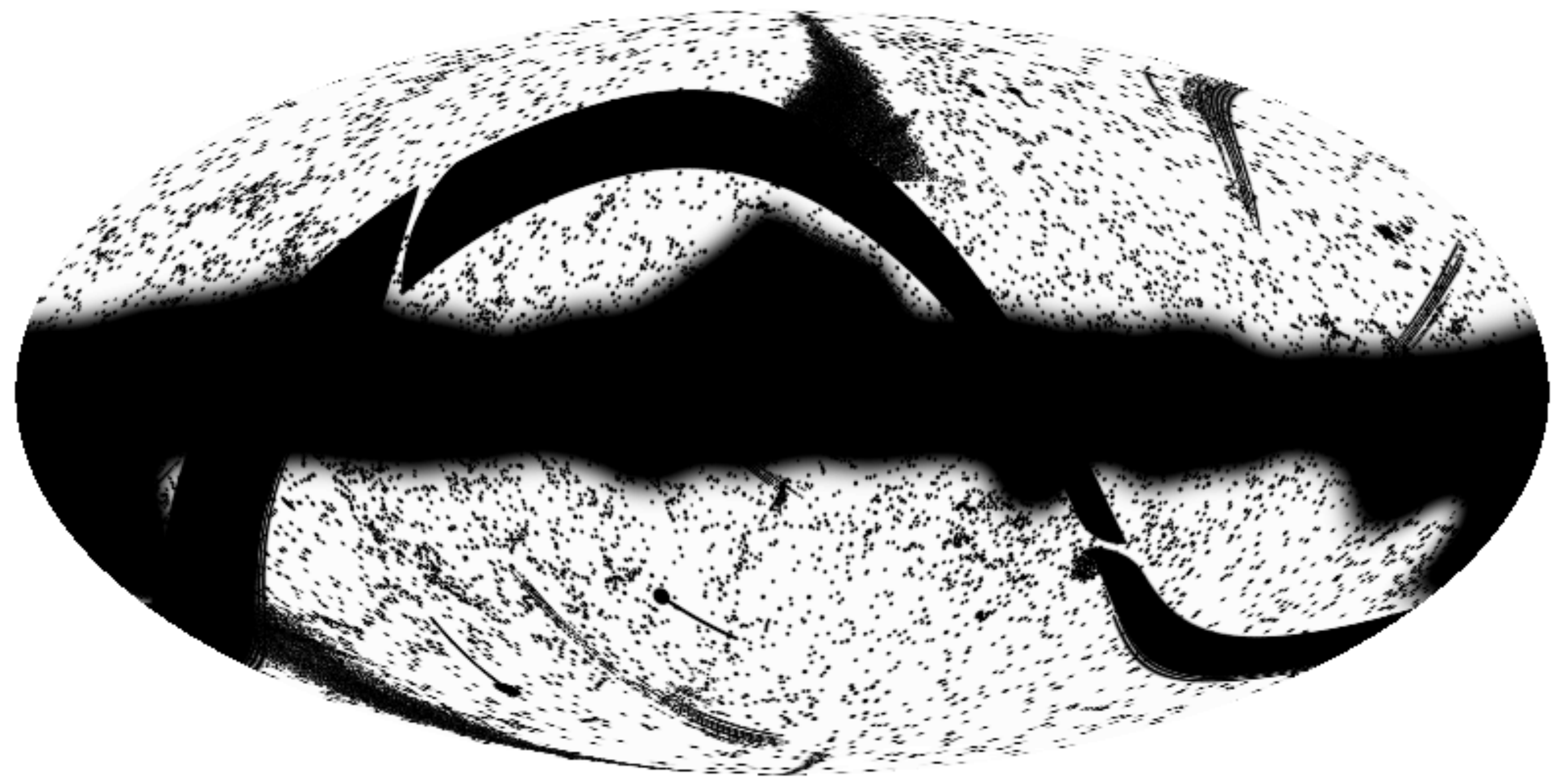}
    \caption{This figure shows the more conservative mask used in the Planck analysis in which foregrounds are modeled, SA24 (left) and the less conservative mask used in our main analysis, SA47 (right).  These
    masks are a product of a point source mask, a mask that excludes pixels observed only in a single survey, and galactic masks with $f_\text{sky}=35\%$ for SA24 and $70\%$ for SA47.  With the SA24 mask, there is $24.3\%$ of the sky available for analysis, with the SA47 mask there is $46.8\%$ of the sky available for analysis. 
\label{fig:masks} }
  \end{center}
\end{figure}

In order to reduce the cosmic variance errors in the power spectrum (and obviate the effects of using a smaller set of cross-spectra), we use more data by analyzing a larger fraction of the sky. This is possible because of Planck's wide frequency coverage which allows a much more aggressive approach to removing foregrounds than was possible in small-scale experiments such as ACT \citep{sievers/etal:2013} and SPT \citep{hou/etal:2012}.  In our analysis, we make use of the publicly available survey maps at 100, 143, 217, 353 and 545 GHz for the power spectrum measurement, and use the 857 GHz data to construct galactic masks.

The galactic masks are obtained by thresholding the 857 GHz map after smoothing with a $\rm{FWHM}=10^\circ$ Gaussian. Our main analysis is based on a galactic mask with $f_\text{sky}=70\%$. To show the robustness of our cleaning procedure, we also present results based on galactic masks with $f_\text{sky}=35\%$, $f_\text{sky}=56\%$, $f_\text{sky}=65\%$, and $f_\text{sky}=75\%$. For comparison with the Planck analysis, we also generate galactic masks with $f_\text{sky}=35\%$ and $f_\text{sky}=56\%$ by thresholding the $353$ GHz map, again after smoothing with $\rm{FWHM}=10^\circ$ Gaussian. All galactic masks are then apodized as described in~\citep{planck:likelihood} by smoothing with $\rm{FWHM}=5^\circ$ Gaussian, subtracting $0.15$ from each pixel, setting pixels with negative values to zero, and dividing the value of each pixel by $0.85$. 

%We begin by constructing sky masks.  By using frequency differences to construct point source masks, we eliminate any biases associated with CMB fluctuations boosting source fluxes.  First, we smooth the 100 and 143 GHz maps to a common resolution and then mask all pixels above $5\sigma$ or below $-5\sigma$ with a 12' mask in the combined $T_{100} - T_{143}$ maps and then repeat the procedure using a CMB-free combination of the 217 and 353 GHz maps. We then multiply by the Planck team's `COM\_PCCS\_SZ-unionMask' to eliminate point sources and mask all pixels that are not observed in both seasons. 

To make our point source masks, we use the masks provided in {\tt HFI\_Mask\_PointSrc\_2048\_R1.10.fits}, which mask point sources detected at $5\,\sigma$ in HFI channels as described in~\citep{planck:component}. For the analysis in which foregrounds are modeled at the level of the power spectrum, as well as for cleaning with 353 GHz, we use the union of these masks from 100 to 353 GHz. For cleaning with 545 GHz and hybrid cleaning we use the union of the masks for 100 to 545 GHz. We apodize these two union masks with a Gaussian with $\rm{FWHM}=30'$. We also mask the pixels that are not observed in both seasons and apodize them with a Gaussian with $\rm{FWHM}=30'$. 

The final masks used in the analysis are the product of galactic, point source, and survey mask. 
Two masks are shown in Figure~\ref{fig:masks}, the more conservative mask leaves $24.3\%$ of the sky for analysis, the less conservative mask leaves $46.8\%$ of the sky for analysis.  We refer to these apodized survey masks based on galactic masks with $f_\text{sky}=35\%$ and $70\%$ as SA24 and SA47. Similarly, we refer to the mask based on the galactic masks with $f_\text{sky}=56\%$, $65\%$ and $75\%$ as SA38, SA44, and SA50, respectively.

To test for stability we have used various different degrees of apodization with negligible effect on cosmological parameters. Since we only use 545 GHz data for cleaning, our point source mask is very conservative. We have also performed analyses in which 545 GHz point sources are not masked and find that they lead to virtually identical cosmological parameters. In addition, we have also estimated cosmological parameters from maps that were filtered to minimize mode coupling and have found no significant shifts. The robustness of the cosmological parameters derived from sky fractions varying from $24\%$ to $50\%$ indicates that CO is not a significant contaminant for our masks. We have also checked this directly using CO maps provided by Planck as well as CO maps made based on the observation that the 100 GHz HFI bandpass includes the CO line while the 94 GHz WMAP W-band does not. 
% {\bf [RF: could also go somewhere in some tiny subsection in 5 where we discuss things one might worry about.]}

\begin{figure} [htbp!]
  \begin{center}
    \includegraphics[totalheight=0.4\columnwidth]{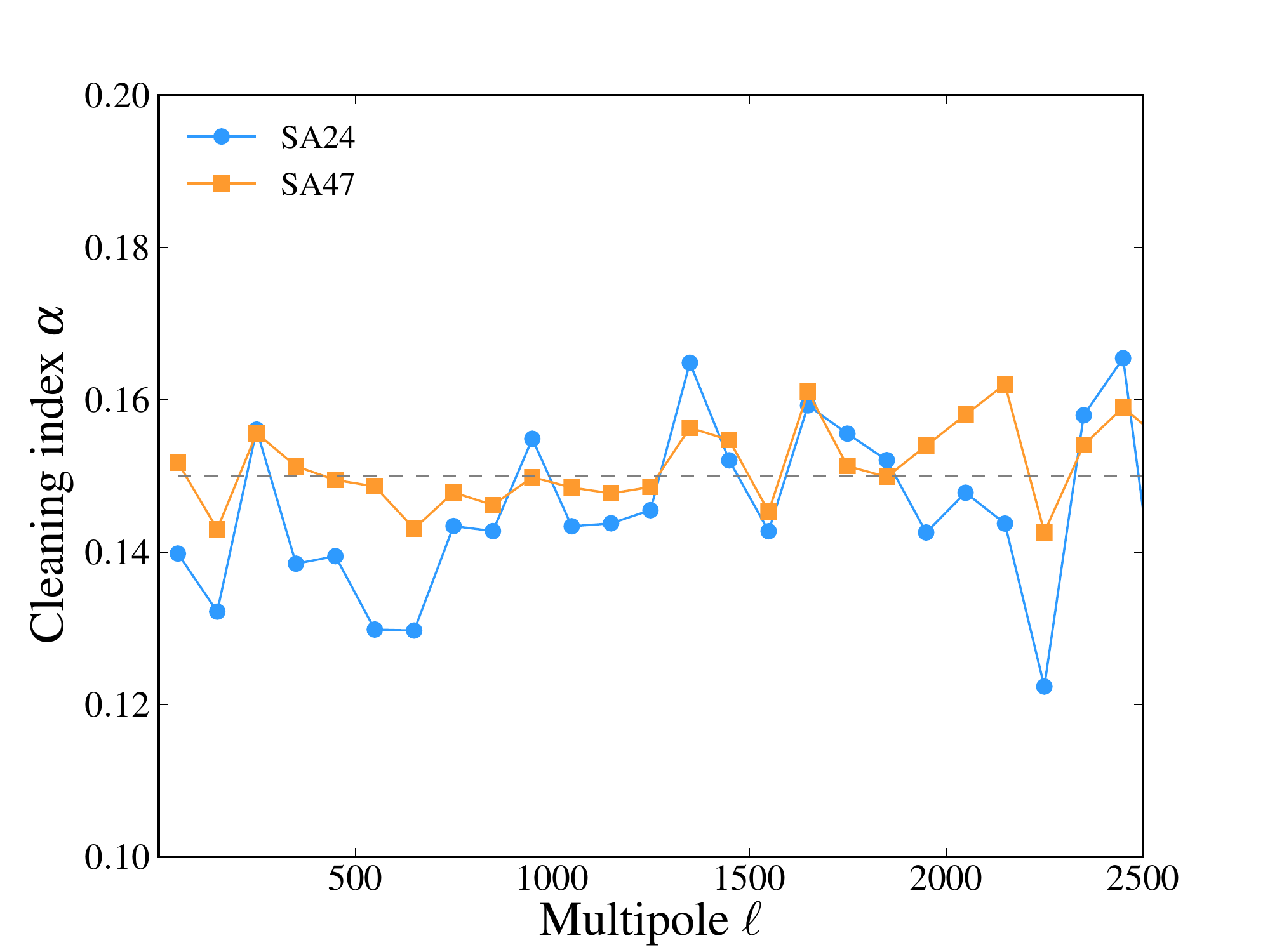}
    \caption{This figure shows the amplitude of the optimal coefficient to minimize foregrounds based on
    minimizing $((1+\alpha)T_{217} - \alpha T_{353})^2$ as a function of multipole number $\ell$ and sky cut. 
    The blue dots are for mask SA24 and the orange squares for SA47.  The dashed line
    shows the value chosen for map cleaning which was fixed to minimize the extragalactic foregrounds. There is no significant variation in the cleaning coefficient with multipole or sky cut.\label{fig:r23}}
  \end{center}
\end{figure}

Using the high frequency maps (343 and 545 GHz), we produce two dust-cleaned maps for each of the intermediate frequency survey maps (100, 143 and 217 GHz). We find the coefficient that minimizes the high $\ell$ angular power spectrum of the clean maps 
\be
T^{clean} _{ij}= (1 +  \alpha_{ij}) T_i -  \alpha_{ij} T_j\,,
\ee
at high galactic latitude. As an example, Figure \ref{fig:r23} shows the coefficient, $\alpha_{217,353}$ that minimizes a linear combination of the 217 and 353 GHz maps and Table \ref{tab:coeff} lists the coefficients used in the map cleaning. For 100 GHz, we extract them from the multipole range $1000<\ell<1200$, for 143 GHz from $1500<\ell<2000$, and for 217 GHz from $2000<\ell<2500$. Figure \ref{fig:r23} shows that the dependence of the cleaning coefficient on multipole is weak so that the choice of $\ell$-range has a very mild effect. Furthermore, the minimum is relatively broad and small changes in cleaning coefficients do not affect the analysis.

\begin{figure*}[htbp!]
\begin{center}
$\begin{array}{@{\hspace{-0.0in}}c}
\includegraphics[width=6.5in]{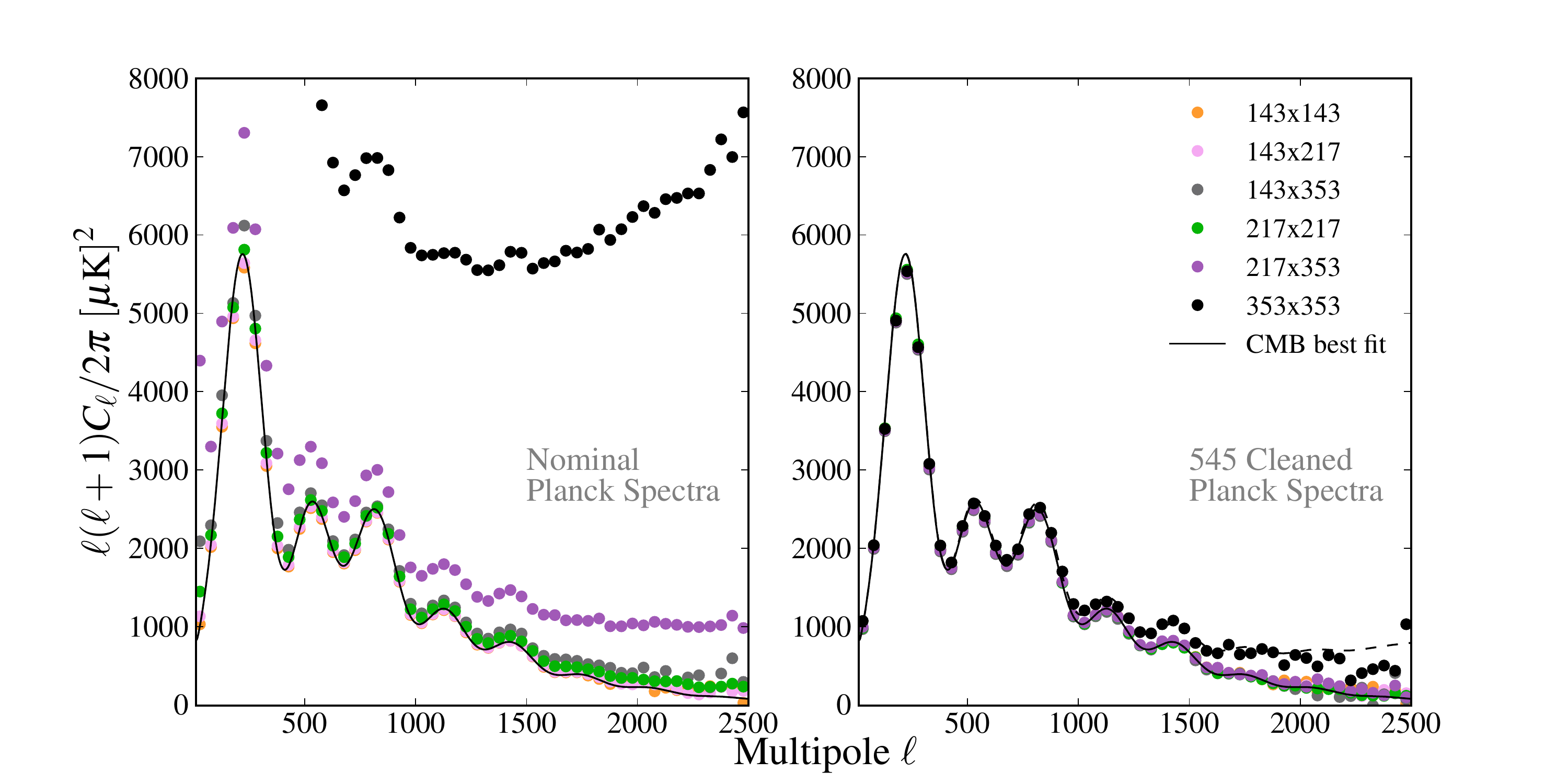}\\[0.0cm]
\end{array}$
\caption{Planck spectra before and after foreground cleaning for mask SA47. \emph{Left panel:} Planck spectra from the uncleaned maps, masked to remove point sources and galactic emission. \emph{Right panel:} Foreground-cleaned Planck spectra using the 545 GHz data to clean the maps used in the cosmological analysis.  The solid line shows the best-fit CMB spectrum while the dashed line indicates the residual Poisson foreground emission after cleaning with an amplitude of $458~(\mu $K$)^2$ at $\ell=2000$, as shown in Figure~\ref{fig:resid}.\label{fig:m3}}
 \end{center}
\end{figure*}

\begin{figure} [ht]
  \begin{center}
    \includegraphics[width=4.3in]{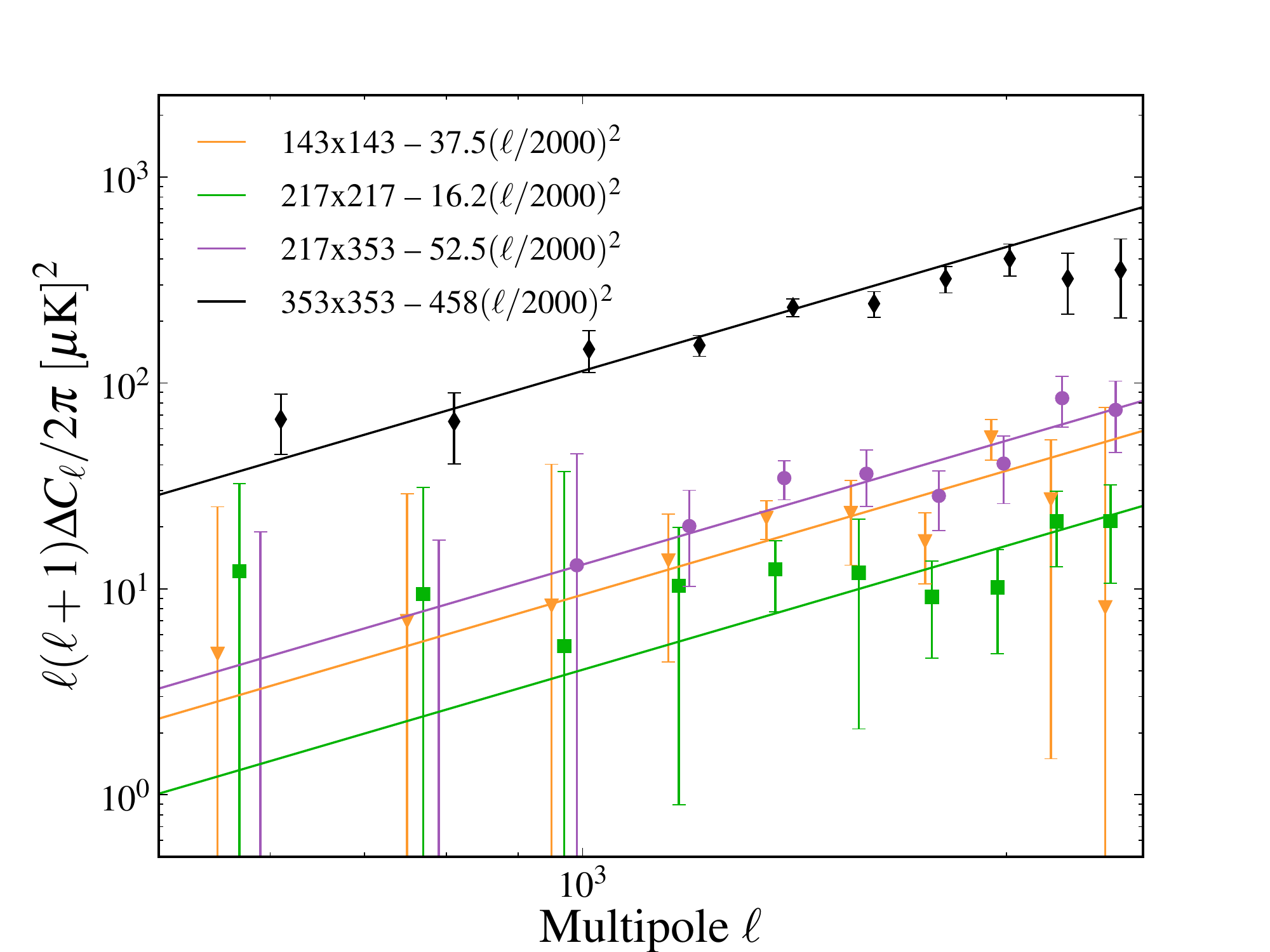}
    \caption{This figure shows the difference between the cleaned power spectra shown in Figure~\ref{fig:m3} and the best-fit CMB spectrum to the 545 GHz-cleaned data.  The solid line shows a fit to this residual of a point source contribution of $458~(\mu $K$)^2$ at $\ell=2000$ for $353 \times  353$ and $52.5~(\mu$K$)^2$ for $217 \times  353$.  If the emission spectra of the dusty galaxies scale as a power law over the 217-545 GHz frequency range, then the residual power in $217 \times  217$ spectra due to dusty galaxies is $4~(\mu$K$)^2$ at $\ell = 2000$.  The fit shown in the plot is for a $16.2~(\mu$K$)^2$ contribution at $217 \times  217$, consistent with the expected additional signal due to synchrotron sources. The fit shown for the $143\times143$ spectrum is $37.5~(\mu$K$)^2.$ \label{fig:resid}}
  \end{center}
\end{figure}

Cleaning with 545 GHz removes the extragalactic dusty sources very effectively. Figure~\ref{fig:m3} compares the `raw' Planck spectra computed for the SA47 sky cut with the 545-cleaned spectra. The figure shows that the 545 cleaning is able to reduce the foreground contribution at 353 GHz by about a factor of $15$ (in power) at $\ell \sim 1800$. Using the double difference technique presented in \citet{planck:likelihood}, we estimate the galactic contribution to the foregrounds at 353 GHz in this $\ell$-range to be around $1000\,\mu K^2$. Cleaning with 545 GHz removes this contribution almost entirely with a residual of $\lesssim10\,\mu K^2$. This implies that more than $90\%$ of the contribution of dusty sources is removed. Figure \ref{fig:resid} shows the level of residual extragalactic point sources based on the difference between the 545-cleaned map power spectrum and our best-fit CMB spectrum.  The plot shows that the residual is well fit by a Poisson term with no evidence for any clustered galaxy term at either 217 GHz or 353 GHz.
\begin{table}[htdp]
\caption{Coefficient $\alpha_{ij}$ used in the cleaning procedure for mask SA47}
\begin{center}
\begin{tabular}{|c||c|c|}
\hline
Template & 353 & 545 \\
\hline
100 & 0.008 &  0.001 \\
143 & 0.03 & 0.002 \\
217 & 0.15 & 0.0085 \\ 
353 & - &  0.064\\
\hline
\end{tabular}
\end{center}
\label{tab:coeff}
\end{table}%
We can use the $217\times353$ and $353\times353$ spectra to estimate the residual level of dusty galaxy point source contributionin the $217\times217$ spectrum. We assume that the flux density for these sources obeys a power law with constant spectral index from 217 GHz through 545 GHz, so that 
\begin{equation}
C^{dust, 217\times353}_\ell = (C^{dust, 217\times217}_\ell C^{dust,353\times353}_\ell)^{1/2} \label{eq:dust_resid}\,.
\end{equation}
This leads to $\ell(\ell+1)C_{2000}^{PS}/2\pi\approx 4\,\mu K^2$ for the $217\times217$ spectrum after subtraction with a 545 GHz template. The difference between the cleaned power spectrum and the CMB-only spectrum that best fits the 545 GHz-cleaned data, is fit by a residual point source amplitude of around $17~(\mu$K$)^2$, consistent with the expected additional contribution from synchrotron sources. For our 353 GHz maps, we can also estimate the residual level of dusty source contamination from Equation~\eqref{eq:dust_resid}. For the $217\times217$ spectrum, we find $\ell(\ell+1)C_{2000}^{PS}/2\pi\lesssim 1\,\mu K^2$, consistent with estimates based on an extrapolation assuming a power law.

Since the spectrum of the galactic dust is close to that of the extragalactic dust, foreground cleaning also significantly reduces the galactic dust contribution in our cleaned maps. We estimate the residual galactic emission using the double difference technique presented in \citep{planck:likelihood}, and find that the residual galactic contribution is less than $\sim 1\,\mu$K$^2$, both for cleaning with 353 and 545 GHz. %\tkRH{[check the number here]}.
While the 353 GHz cleaning is slightly more effective at removing both the extragalactic clustered dust contribution and the galactic foregrounds, the cost of this cleaning procedure is a 25\% increase in the noise in the cleaned maps. 
%With the 353 GHz cleaning, we find that the 100, 143 and 217 GHz power spectra are very consistent between the SA24 and SA47 masks. 

For our analysis, we compute the cross-spectra of the 353 GHz-cleaned maps with the 545 GHz-cleaned maps. This hybrid approach has the advantage of nearly halving the additional noise cost associated with using the high-frequency maps and being insensitive to the details of the 353 GHz and 545 GHz maps. We will focus on results from the hybrid cleaned spectra using the SA47 sky cut, but we have also computed cosmological parameters for a number of other sky cuts as well as for 353 GHz and 545 GHz cleaned spectra to check for stability.  We show a subset of the various approaches in Figure~\ref{fig:nsom}.

\section{Cosmological parameters using cleaned spectra}
\label{sec:parameters}
\begin{figure*}[htbp!]
\begin{center}
$\begin{array}{@{\hspace{-0.in}}l@{\hspace{-0.25in}}l}
\includegraphics[width=3.6in]{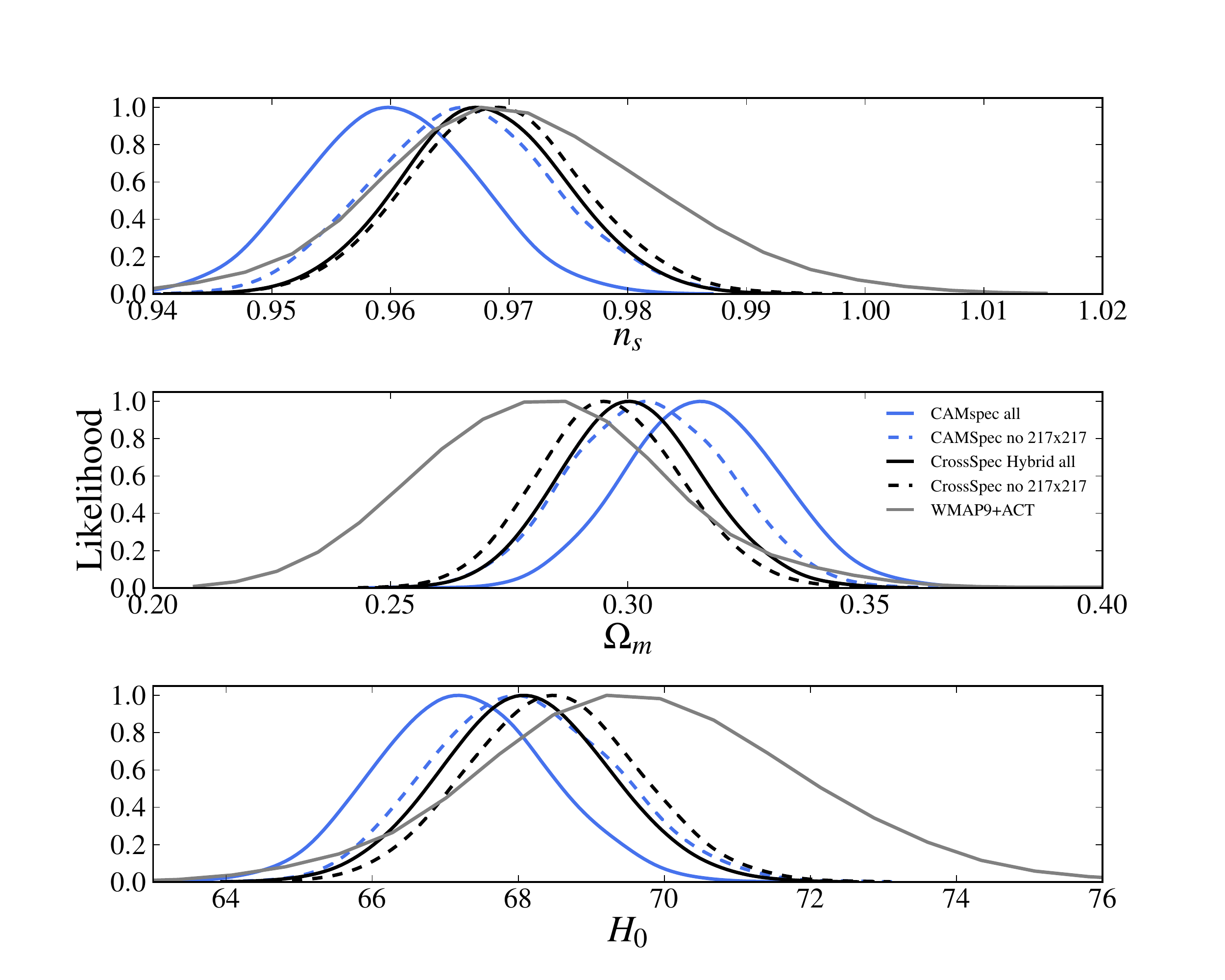}&
\includegraphics[width=3.6in]{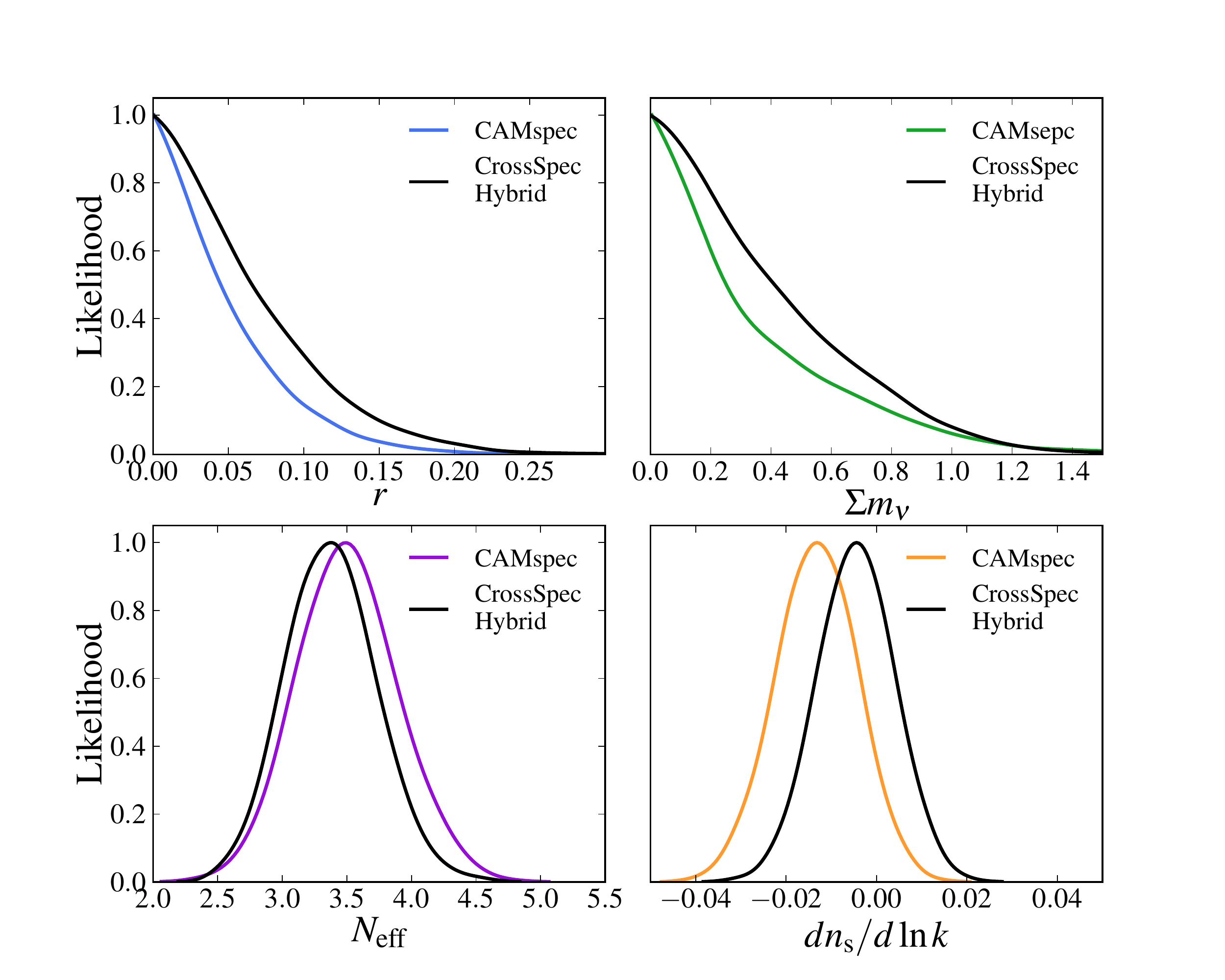}\\[0.0cm]
\end{array}$
\caption{Constraints on key parameters in the $\Lambda$CDM model and extensions in the presence of high-frequency foreground cleaning. \emph{Left panel:} The top, middle and bottom panels show the marginalized one-dimensional likelihoods for the scalar spectral index, $n_s$, the matter density $\Omega_m$ and the Hubble constant $H_0$ in the $\Lambda$CDM model. The solid blue line is the standard Planck result for the CAMSpec likelihood including the $100\times100$, $143\times 143$, $217\times 217$ and $143\times 217$ spectra. The dashed blue line shows the results when the $217\times 217$ spectra are not used; these correspond to results presented in Figure B3 of \citep{planck:parameters}. The solid and dashed black lines show the same for the cleaned spectra presented here. \emph{Right panel}: Constraints on the tensor-to-scalar ratio (top left panel) and mass of the neutrino (top right panel) are weakened with cleaning of the spectra, and $N_\mathrm{eff}$ (bottom right hand panel) shifts to values slightly more consistent with three neutrino species. The cleaned spectra do not show a preference for running of the scalar spectral index. \label{fig:cleaned}}
\end{center}
\end{figure*}

For the estimate of cosmological parameters, we use Planck's `clik' likelihood code. We leave the Commander and lowlike likelihood unchanged, but modify CAMspec to accomodate the changes in our analysis.  Our basic approach follows the procedures outlined in Planck papers XV and XVI as closely as possible.  We replace the Planck team's $100\times100$, $143\times143$, $143\times217$ and $217\times217$ spectra with the `hybrid cleaned' spectra that we have computed from the publicly available maps using the procedures described in Section~\ref{sec:clean}.  We compute the covariance matrices for the spectra using the same approximations described in Appendix A of Planck paper XV with appropriate modifications to take into account the noise in the 353 GHz and 545 GHz maps used in the cleaning. Since the noise is colored, noise corrections are required for an accurate covariance matrix. We measure these corrections by fitting the Planck noise model to noise spectra obtained from survey differences. As fiducial models for the covariance matrices, we use the best-fit from a run based on a covariance matrix for the same masks and cleaning procedure in which they were taken to be CMB only. We retain the modeling of extra-galactic foregrounds used in CAMspec and thus the nuisance parameters~\citep{planck:likelihood} and use the conventions and units given there.  

We then use CAMB~\citep{Lewis:1999bs} and CosmoMC~\citep{Lewis:2002ah} to derive constraints on cosmological parameters for the Planck+WP data set with our likelihood. We find best-fit cosmological parameters that are closer to pre-Planck values than the results in~\citep{planck:parameters}.
Figure~\ref{fig:cleaned} shows the marginalized one-dimensional likelihoods for the three parameters in the standard $\Lambda$CDM model which showed significant shifts in the recent Planck results relative to previous analyses. The constraints using the cleaned spectra (indicated by the blue lines in Figure~\ref{fig:cleaned}) are more consistent with previous results, such as those from ACT~\citep{sievers/etal:2013}, shown in gray.  
%By using spectra computed from the cleaned season crosses, the power spectrum amplitude is higher for $\ell > 1500$. This increased amplitude leads to a shift in cosmological parameters.  
The most notable shifts are along a modest degeneracy line between $n_s, H_0$ and $\Omega_m$:\footnote{In all comparisons, we will use the standard deviation of the experiment with the larger uncertainty when stating that one measurement is within $x\,\sigma$ of the other.}
\begin{table}[htdp]
\caption{Constraints on the $\Lambda$CDM model parameters in the cleaning procedure compared to the nominal Planck parameters.}
\begin{center}
\begin{tabular}{|c|c|c|}
\hline
Parameter & Planck constraint & Hybrid cleaning constraint $(f_{sky}=0.47)$\\
\hline
\hline
 \multicolumn{3}{|c|}{Primary Parameters  } \\
\hline
$\Omega_bh^2$ & $0.02204 \pm 0.00028$&$ 0.02197 \pm 0.00026$\\
$\Omega_ch^2$& $0.1199 \pm 0.0026$& $0.1170 \pm 0.0025$\\
$\theta_A$&$1.04131 \pm 0.00062$ &$ 1.04066 \pm 0.00056$\\
$\tau$ & $0.089\pm0.013$&$0.090\pm 0.013$\\
$n_s$ & $0.9604\pm 0.0073$&$0.9686\pm 0.0069$\\
$\ln(10^{10}A_s)$&$3.088\pm 0.025$&$3.082\pm 0.025$\\
\hline
 \multicolumn{3}{|c|}{Derived Parameters  } \\
 \hline
$H_0$&$67.3\pm 1.2$&$ 68.0\pm 1.1$\\
$\Omega_m$&$0.316 \pm 0.016$&$ 0.302\pm  0.015$\\
$\sigma_8$&$0.829 \pm 0.016$&$ 0.818\pm  0.012$\\
\hline
 \multicolumn{3}{|c|}{Secondary Parameters  } \\
 \hline
$A_{100}^\mathrm{PS}$& $171 \pm60$ &$170\pm43$\\
$A_{143}^\mathrm{ PS}$& $54\pm 13$ &$59\pm13$\\
$A_{217}^\mathrm{ PS}$& $107 \pm 17$ &$17.5\pm6.6$\\
$A_{143}^\mathrm{ CIB}$& $8.1\pm 5.4$ &$7.5\pm5.1$\\
$A_{217}^\mathrm{ CIB}$& $28.9 \pm 7.4$ &$1.6\pm1.5$\\  
$A^\mathrm{ tSZ}$& $5.2\pm2.8$ &$4.5\pm1.8$\\
$r^\mathrm{ PS}_\mathrm{ 143x217}$& $0.883 \pm 0.079$ &$0.94\pm0.055$\\
$r^\mathrm{ CIB}_\mathrm{143x217}$& $0.429 \pm 0.221$ &$0.71\pm0.25$\\
$\gamma_{CIB}$& $0.53\pm 0.13$ &$0.81\pm0.19$\\
$c_{ 100}$& $1.00058\pm 0.00040$ &$1.00259\pm0.00044$\\
$c_{ 217}$& $0.99638\pm0.0014$ &$0.99591\pm0.00083$\\
$\xi_\mathrm{ tSZxCIB}$& $0.47 \pm 0.29$ &$0.54\pm0.29$\\
$A_\mathrm{ kSZ}$& $4.5\pm2.9$ &$4.0\pm2.7$\\
\hline
\end{tabular}
\end{center}
\label{table:cosmo_params}
\end{table}%

\begin{figure}[htbp!]
\begin{center}
$\begin{array}{@{\hspace{-0.2in}}l}
\includegraphics[width=6.3in]{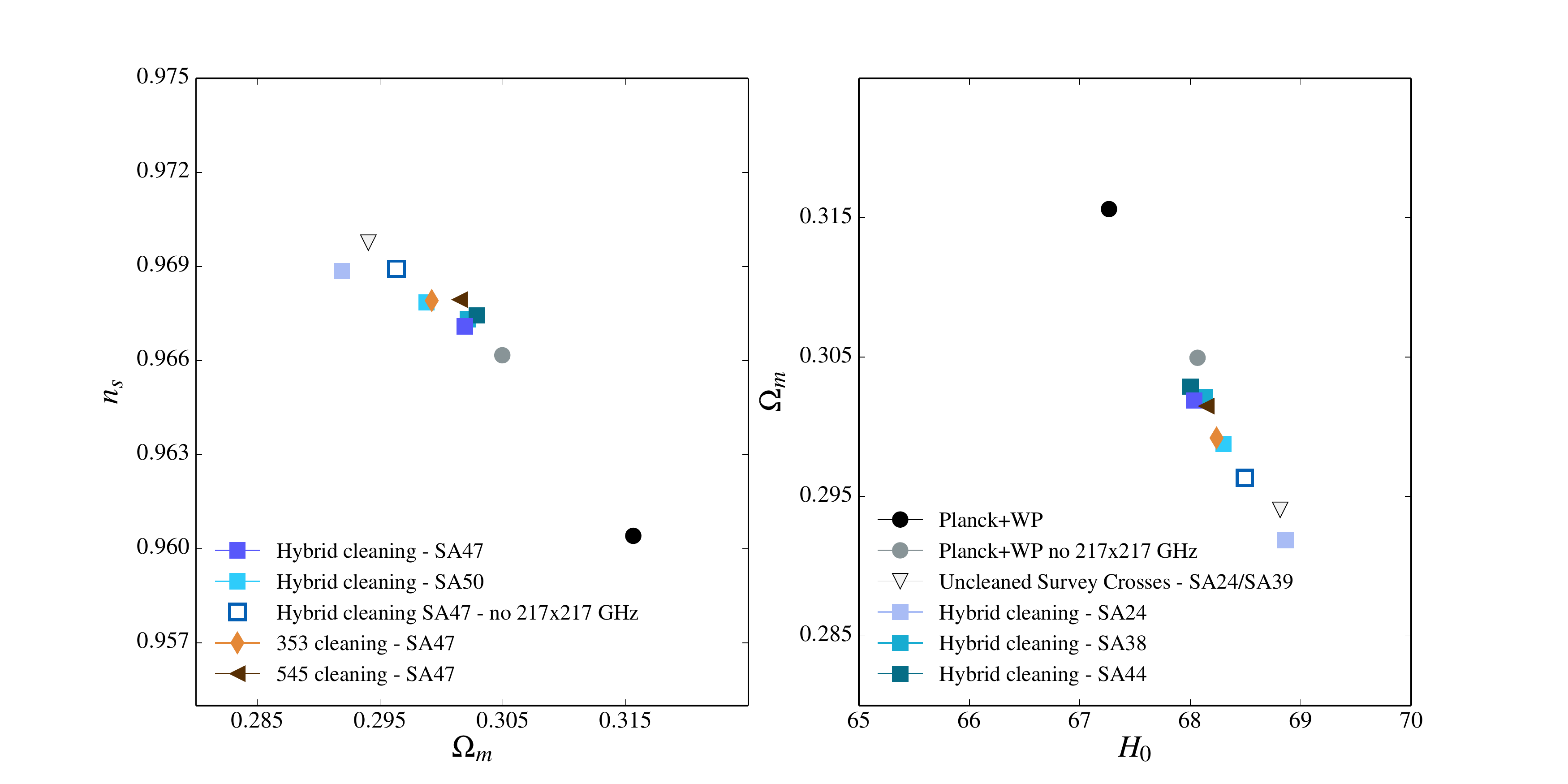}\\ [0.0cm]
\end{array}$
\caption{Constraints in the $\Omega_m-n_s$ plane (left) and $H_0-\Omega_m$ plane (right) for various cleaning strategies and datasets, compared to the Planck results. The circles show results obtained with the nominal Planck data, the squares show results from the hybrid cleaning procedure, the diamond are obtained when only cleaning with the 353 GHz data, and the triangle when using the 545 GHz data to clean the lower frequencies. For comparison, the results are shown for the season cross-spectra without additional cleaning (upside-down triangle). The cosmological results are robust to a change in the cleaning procedure. The uncleaned survey crosses use the combination of masks used by Planck, modified to mask pixels observed in only one survey. We refer to them as SA39 for 100 GHz and SA24 for 143 and 217 GHz. \label{fig:nsom}}
 \end{center}
\end{figure}

\begin{figure}[htbp!]
\begin{center}
$\begin{array}{@{\hspace{-0.2in}}l}
\includegraphics[width=3.5in]{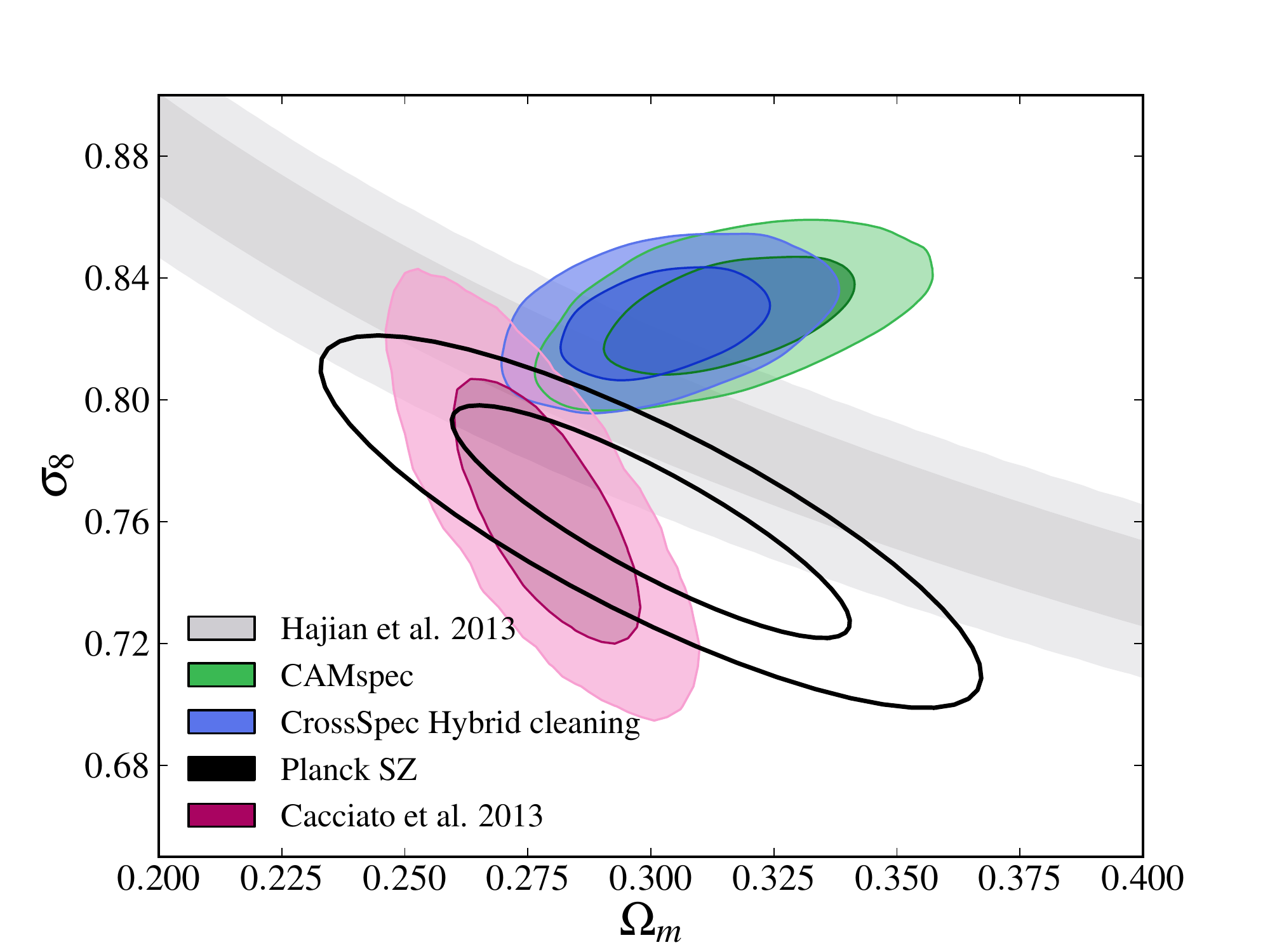}\\ [0.0cm]
\end{array}$
\caption{Constraints on $\sigma_8$ for the hybrid cleaning method (in blue), in comparison to the constraints from the Planck CMB constraints (green filled contours \citep{planck:parameters}), Planck SZ clusters (black unfilled contours~\citep{planck:SZ}), constraints on the SZ emission from cross correlation of CMB maps and galaxy cluster catalogs (grey band~\citep{hajian/etal:2013}) and from a combination of galaxy clustering and lensing (pink filled contours~\citep{cacciato/etal:2013}).  \label{fig:sigma8_om}}
 \end{center}
\end{figure}

\begin{itemize}
\item \textbf{the spectral index} $n_s$, is about $1\,\sigma$ higher in our analysis with mean $0.9686\pm 0.0069$, still significantly different from $n_s = 1$.
\item \textbf{the Hubble constant} $H_0 = 68.0\pm 1.1$ $\mathrm{km}~\mathrm{s}^{-1}\mathrm{Mpc}^{-1}$, shifts upwards from the best-fit Planck value by $0.6\,\sigma$, within $1\,\sigma$ of the value inferred from the WMAP9 data, $H_0 = 70.0 \pm 2.2$ $\mathrm{km}~\mathrm{s}^{-1}\mathrm{Mpc}^{-1}$ and within $2\,\sigma$ of published local estimates of the Hubble constant, $73.8 \pm 2.4\,\mathrm{km}~\mathrm{s}^{-1}\mathrm{Mpc}^{-1}$~\citep{riess2011} and $74.3 \pm 1.5 \pm 2.2\,\mathrm{km}~\mathrm{s}^{-1}\mathrm{Mpc}^{-1}$~\citep{freedman2012}.  Our recomputed value is even closer to the  recent re-evaluation based on a new distance to NGC 4258,   $70.6 \pm 3.3$ $\mathrm{km}~\mathrm{s}^{-1}\mathrm{Mpc}^{-1}$~\citep{efstathiou2013}, a difference of only $0.8\,\sigma$ and within $1.8\,\sigma$ of the analysis based on NGC 4258, Cepheids in the LMC and the Milky Way presented there which yielded $72.5 \pm 2.5$ $\mathrm{km}~\mathrm{s}^{-1}\mathrm{Mpc}^{-1}$.
Our value for the Hubble constant is also more consistent with gravitational lensing timing measurements~\citep{suyu/etal:2013}. Of course, it should be kept in mind that even after a shift by $0.6$ Planck $\sigma$ the CMB in the context of $\Lambda$CDM still indicates a lower value of $H_0$ than the astonomical measurements. 
\item \textbf{the matter density} $\Omega_m$ \textbf{and} $\sigma_8$ are about $1 \sigma$ lower. We find $\Omega_m=0.302\pm  0.015$ compared to Planck's value of $\Omega_m=0.316$, and $\sigma_8=0.818\pm0.012$ compared to $\sigma_8=0.829$ reported by Planck. The lower matter density and $\sigma_8$ also imply a lower number of rich clusters which determine a combination that scales as $\sigma_8 (\Omega_m/0.27)^{0.3}$.  The Planck team reports an amplitude of $0.87 \pm 0.02$ \citep{planck:parameters}, $\sim 3 \sigma$ discrepant from the value inferred from the Planck SZ cluster measurement: $0.79 \pm 0.01$~\citep{planck:clusters}. Our analysis leads to the constraint $\sigma_8 (\Omega_m/0.27)^{0.3}=0.84\pm0.02$ lowering the tension to $\sim 2 \sigma$, but not eliminating it entirely. Measurements of the X-ray cluster-SZ cross-correlation yield the constraint $\sigma_8 (\Omega_m/0.3)^{0.26} = 0.797 \pm 0.015$~\citep{hajian/etal:2013}. We find $\sigma_8 (\Omega_m/0.3)^{0.26} = 0.818 \pm 0.019$, a difference of only $1\,\sigma$. The normalization of the local mass function determined from X-ray cluster data yields $\sigma_8 (\Omega_m/0.25)^{0.47}=0.813\pm0.013$~\citep{vikhlinin/etal:2009}. We find $\sigma_8 (\Omega_m/0.25)^{0.47}=0.892\pm0.028$, still in tension with this measurement at $~2\,\sigma$ after accounting for systematics mentioned in~\citep{vikhlinin/etal:2009}. The CFHTLens analysis finds $\sigma_8 = 0.799 \pm 0.015$, within $1\,\sigma$ of our measurement, but finds $\Omega_m = 0.27 \pm 0.010$ about $2\,\sigma$ lower than our measurement. Some of these results are summarized in Figure~\ref{fig:sigma8_om}. 
\end{itemize}
For the cleaned spectra, the best-fit values of the foreground parameters change, as shown in Table~\ref{table:cosmo_params}. In general we expect the amplitudes of foreground parameters such as the point sources amplitudes at 217 GHz to reduce dramatically, where the amplitudes of the CIB at 143 and 217 GHz are consistent with a non-detection. 
The amplitude of the CIB sources at 217 GHz is also reduced from $A^\mathrm{CIB}_{217} = 27~(\mu\mathrm{K})^2$ to  $A^\mathrm{CIB}_{217} < 10~(\mu\mathrm{K})^2$ at 95\% confidence, and the upper limit at 143 GHz also slightly reduces in amplitude. 

%The stringent masking of SZ clusters also removes much of the SZ signal at 143 GHz, but does not improve the constraint on the correlation between the CIB and tSZ signals, partly because the amplitude of the tSZ signal itself is so low. As might be expected, the cleaning of the spectrum does not change the inter-spectra calibration numbers as the cleaning removes residual power at high multipoles rather than at low-$\ell$. 

In the Planck team analysis \citep{planck:parameters}, the cosmological parameters change when the analysis excludes the $217 \times 217$ spectrum, as shown in Figure~\ref{fig:cleaned}. To a smaller extent this trend continues even when using cleaned data. The shifts are now roughly consistent with expectations based on simulations, but they could nevertheless indicate a remaining systematic in the $217\times217$ data that is not removed in the cleaned survey cross-spectra. We note that the best-fit parameters obtained in our hybrid cleaning analysis are very close to the best-fit values for the Planck team analysis without the $217 \times 217$ power spectrum. 
\begin{figure}[htbp!]
\begin{center}
$\begin{array}{@{\hspace{-0.2in}}l}
\includegraphics[width=3.8in]{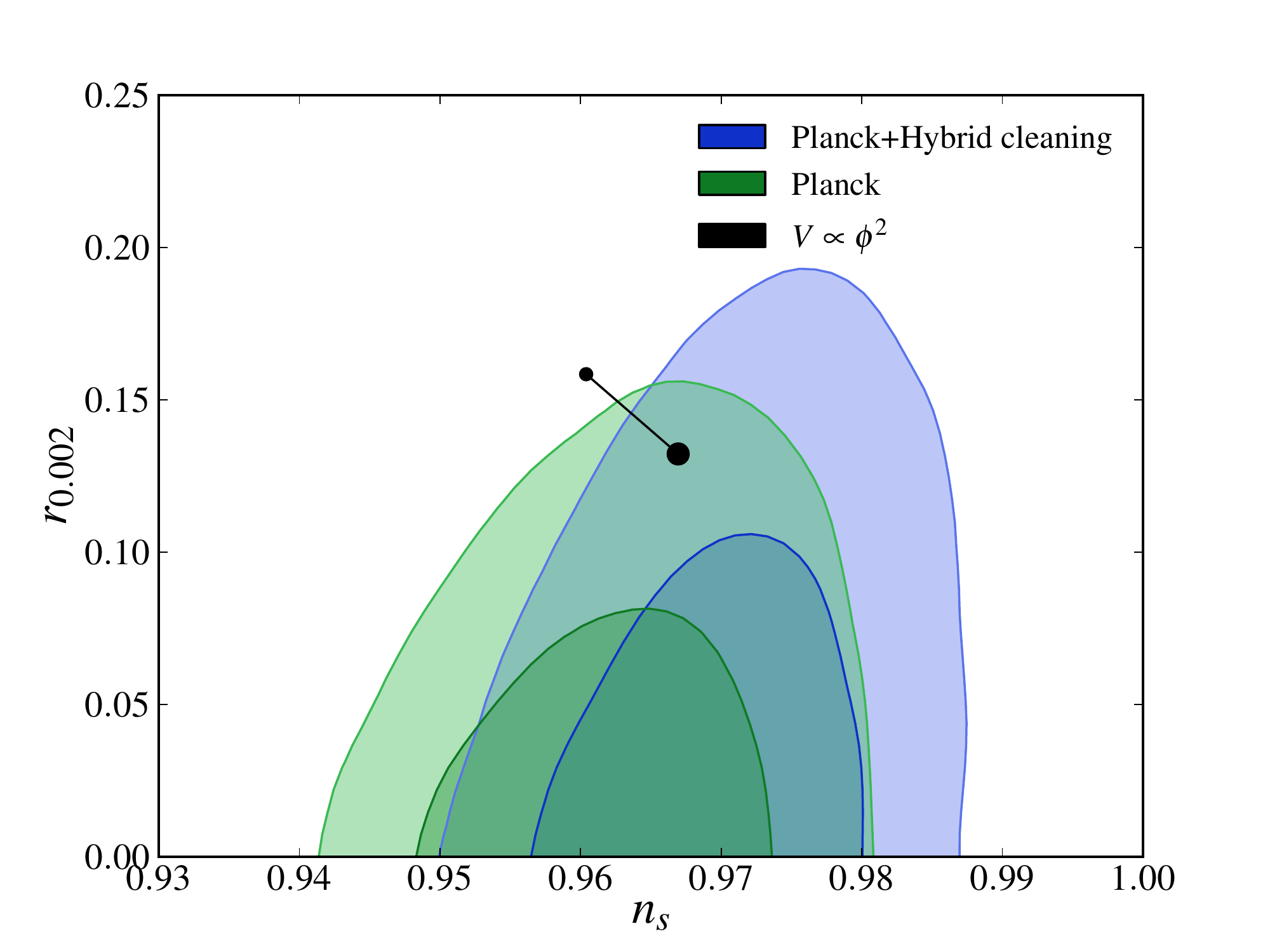}\\ [0.0cm]
\end{array}$
\caption{Constraints on $n_s-r$ plane.  The black dots indicated the predicted values for $m^2 \phi^2$ inflation with 50 and 60 e-folds
between reheating scale and the horizon scale today.  \label{fig:n_s-r}}
\end{center}
\end{figure}

Using the 353 GHz and 545 GHz hybrid cleaning on the publicly available maps, we have also recomputed the best-fit parameters for several extensions of the standard $\Lambda$CDM model. Our re-analysis, shown in Figure \ref{fig:cleaned}, finds no evidence for new physics in contrast with~\citep{hou/etal:2012}: the mean of the posterior for the effective number of relativistic species is $3.34 \pm 0.35$, within $\sim1~\sigma$ of the standard value, and a best-fit point of $N_\text{eff}=3.13$. In addition, there is no evidence for neutrino mass, and the best-fit does not provide evidence for the running of the spectral index, $dn_s/d\log(k) =  -0.0046 \pm 0.0085$.  
Figure~\ref{fig:n_s-r} shows the recomputed constraints on the $n_s-r$ plane: the $m^2 \phi^2$ model lies within the $95\%$ confidence region; however, the data favors models with lower values of $r$ and larger values of $n_s$.
\section{Why Does This Analysis Yield Different Cosmological Parameters?}
\label{sec:compare}
Our analysis of the publicly available Planck data has led to parameters that are different from those found by the Planck team~\citep{planck:parameters} by about $1\,\sigma$. In this section we summarize what might be responsible for them. 
\subsection{Not Due to the Large $f_{sky}$}
\begin{table*} [htbp!]
\caption{Table of mean values of parameters from hybrid cleaning with different values of $f_{sky}.$ \label{table:fsky}}
\begin{center}
\begin{tabular}{|c|c|c|c|c|}
\hline
& $f_{sky}=0.38$& $f_{sky}=0.44$&$f_{sky}=0.47$&$f_{sky}=0.50$\\
\hline
\hline
$\Omega_c h^2$          &0.1174& 0.1173 &0.1170 &0.1167\\
$n_s$                   &0.9681& 0.9683 &0.9685 &0.9683\\
$h$                     &0.681 & 0.680  &0.680  &0.682 \\
 100 $\Omega_b h^2$     &2.197 & 2.197  &2.197  &2.204 \\
 $\log(10^{10} A_s)$    &3.087 & 3.082  &3.082  &3.082 \\
 $\tau $                &0.092 & 0.090  &0.090  &0.091 \\
 \hline
 $-2\ln\mathcal{L}_\text{CAMspec} $ &7529.96&7530.04&7542.45&7626.79\\
 $-2\ln\mathcal{L}_\text{Commander} $ &-8.14&-8.31  &-8.32   &-8.22     \\
 $-2\ln\mathcal{L}_\text{lowlike} $ &2014.60&2014.60&2014.57&2014.70\\
 \hline
\end{tabular}
\end{center}
\end{table*}%
We work with survey cross-spectra rather than the detector set spectra used by the Planck team. As discussed, this means we have less data available for the same $f_\text{sky}$. Because we are cleaning the maps, we can easily increase the fraction of the sky used in the analysis to use more data. This increase in $f_\text{sky}$ may have led to shifts in cosmological parameters. Here we show that the computed power spectra and the best-fit parameters are insensitive to the fraction of the sky used for the analysis. There is no statistically significant difference between the SA24 and SA47 power spectra even when binned with $\Delta \ell = 200$. Furthermore, as shown in Figure \ref{fig:nsom} and in Table \ref{table:fsky}, the cleaned cross-season power spectra and best-fit parameters are consistent for a large range of sky cuts. This suggests that there is minimal residual galactic contribution after cleaning the spectra with the 353 GHz and 545 GHz maps. This is confirmed by measurements of the residual galactic contribution based on the double difference method described in~\citep{planck:likelihood}. 

Table~\ref{table:fsky} also justifies the choice of sky fraction used in our main analysis. On the one hand we would like to use as much of the sky as possible. On the other hand we know that our foreground model will eventually break down. We interpret the sharp increase in $-2\ln\mathcal{L}_\text{CAMspec}$ between the third and fourth column as a signal that this occurs for $f_\text{sky}>47\%$. The table shows that $-2\ln\mathcal{L}_\text{CAMspec}$ is stable for galactic masks with $f_\text{sky}\leq 47\%$ so that our analysis is based on the mask with the largest $f_\text{sky}$ for which we know that the foregrounds are well modeled. Notice, however, that the cosmological parameters do not yet shift significantly for the mask with $f_\text{sky}= 50\%$ corresponding to a galactic mask with $f_\text{sky}=75\%$.

\subsection{Not Due to the Cleaning Procedure}

In the Planck team analysis foregrounds are modeled at level of the power spectrum and the amplitudes of the various templates are marginalized over, with the exception of diffuse galactic emission, for which a template is subtracted from the measured spectra. In our main analysis we clean the maps, which allows us to use significantly larger fractions of the sky. This cleaning could itself introduce systematics and be responsible for shifts in cosmological parameters. To minimize such systematics, our main analysis relies on a hybrid cleaning method based on two sets of maps cleaned with 353 GHz and 545 GHz so that no $353\times353$ or $545\times 545$ spectra are used. In table~\ref{table:cleaning}, we compare cosmological parameters derived from this hybrid cleaning as well as cleaning with 353 GHz and 545 GHz only, all based on the mask SA47. We see that the different ways of cleaning the maps lead to very consistent results, suggesting that our cleaning does not introduce systematics that lead to significant shifts in cosmological parameters. For comparison, we also present an analysis applying Planck's foreground model to season cross-spectra. For these uncleaned survey crosses, we use the combination of masks used by Planck, modified to mask pixels observed in only one survey. We refer to them as SA39 for 100 GHz and SA24 for 143 and 217 GHz. As shown in Figure~\ref{fig:nsom}, for similar $f_\text{sky}$ our cleaning also leads to results consistent with those found by applying Planck's foreground model to season cross-spectra. 
\begin{table*} [htbp!]
\caption{Table of mean values of parameters based on various cleaning procedures. \label{table:cleaning}}
\begin{center}
\begin{tabular}{|c|c|c|c|}
\hline
& $353$ GHz& $545$ GHz&Hybrid cleaning \\
\hline
\hline
$\Omega_c h^2$       &0.1165 &0.1171 &0.1170 \\
$n_s$                &0.9681 &0.9675 &0.9685 \\
$h$                  &0.682 &0.680 &0.680 \\
 100 $\Omega_b h^2$  &2.202 &2.200 &2.197 \\
 $\log(10^{10} A_s)$ &3.080 &3.084 & 3.082 \\
 $\tau $             &0.089 &0.091 &0.090 \\
 \hline
 $-2\ln\mathcal{L}_\text{CAMspec} $ &7590.66&7560.26&7542.46\\
 $-2\ln\mathcal{L}_\text{Commander} $ &-8.22&-7.79  &-8.33  \\
 $-2\ln\mathcal{L}_\text{lowlike} $ &2014.66&2014.66&2014.57\\
 \hline
\end{tabular}
\end{center}
\end{table*}%

\begin{figure}[htbp!]
\begin{center}
$\begin{array}{@{\hspace{-0.1in}}l}
 \includegraphics[totalheight=0.45\textheight]{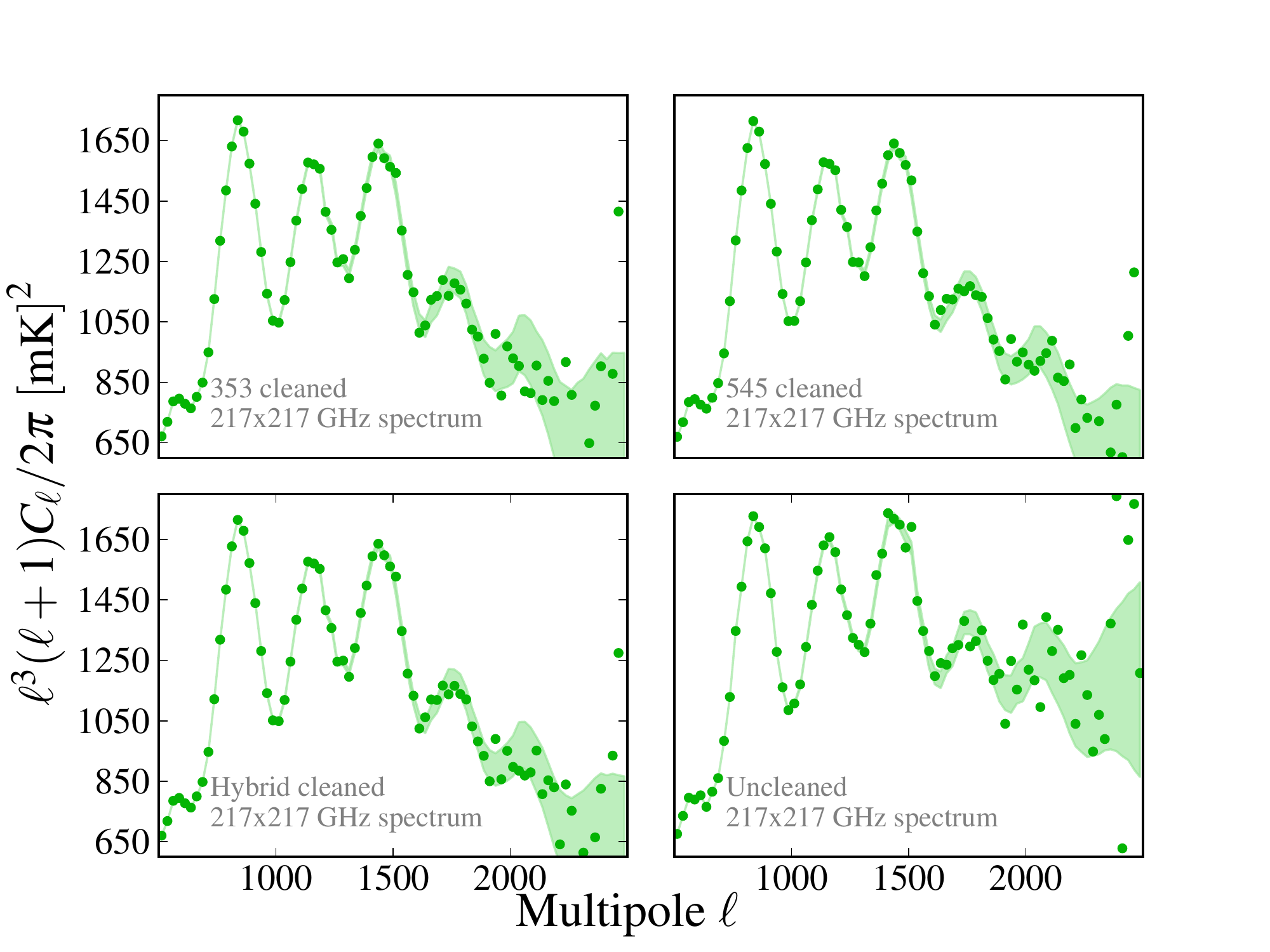}\\ [0.0cm]
\end{array}$
    \caption{Predicted versus observed $217\times217$ spectrum. The shaded bands show the predicted envelope for the $217\times217$ spectrum from the $100\times100$, $143\times143$ and $143\times217$ spectra. The points show the observed $217\times217$ spectrum in each case. By comparing the predicted spectrum from the cleaning done using the 353 GHz spectrum, we see that the data are consistent with the prediction in the $\ell> 1500$ region. The same is true for the hybrid cleaning procedure. The prediction is based on equation~\eqref{eq:predict} and the best-fit foreground spectrum.\label{fig:predict}}
  \end{center}
\end{figure}

To check for potential systematics at the level of the power spectra, we use Equation~\ref{eq:predict} to predict the $217\times 217$ spectrum based on the spectra at other frequencies for the different approaches we used. The result is shown in Figure~\ref{fig:predict}. The trend seen in Figure~\ref{fig:consistency} with observed data systematically below the predicted band for $1700<\ell <1900$ and above the band for the range $2100<\ell< 2400$ is not seen in any of the spectra computed from season cross-spectra. In addition, the panels which show the cleaning procedures highlight the reduction in the error bars due to increased $f_\text{sky}$ possible when applying the cleaning procedure. 

Finally, we test that the cleaning is robust to removing the CIB modeling. From the full set of parameters presented in Table~\ref{table:cosmo_params}, we note that the CIB amplitude at 217 GHz are consistent with zero in the cleaned spectra. We find that the cosmological parameters do not shift by more than $0.2\,\sigma$ when removing the CIB terms completely and setting $A^{\mathrm{CIB}}_{217}$ and $\gamma_{\mathrm{CIB}}$ to zero.

\subsection{Mostly due to the  Detector Set Spectra or Regions Observed Only Once}
\label{sec:observing}
\begin{figure*}[htbp!]
\begin{center}
$\begin{array}{@{\hspace{-0.4in}}l}
 \includegraphics[totalheight=0.5\textheight]{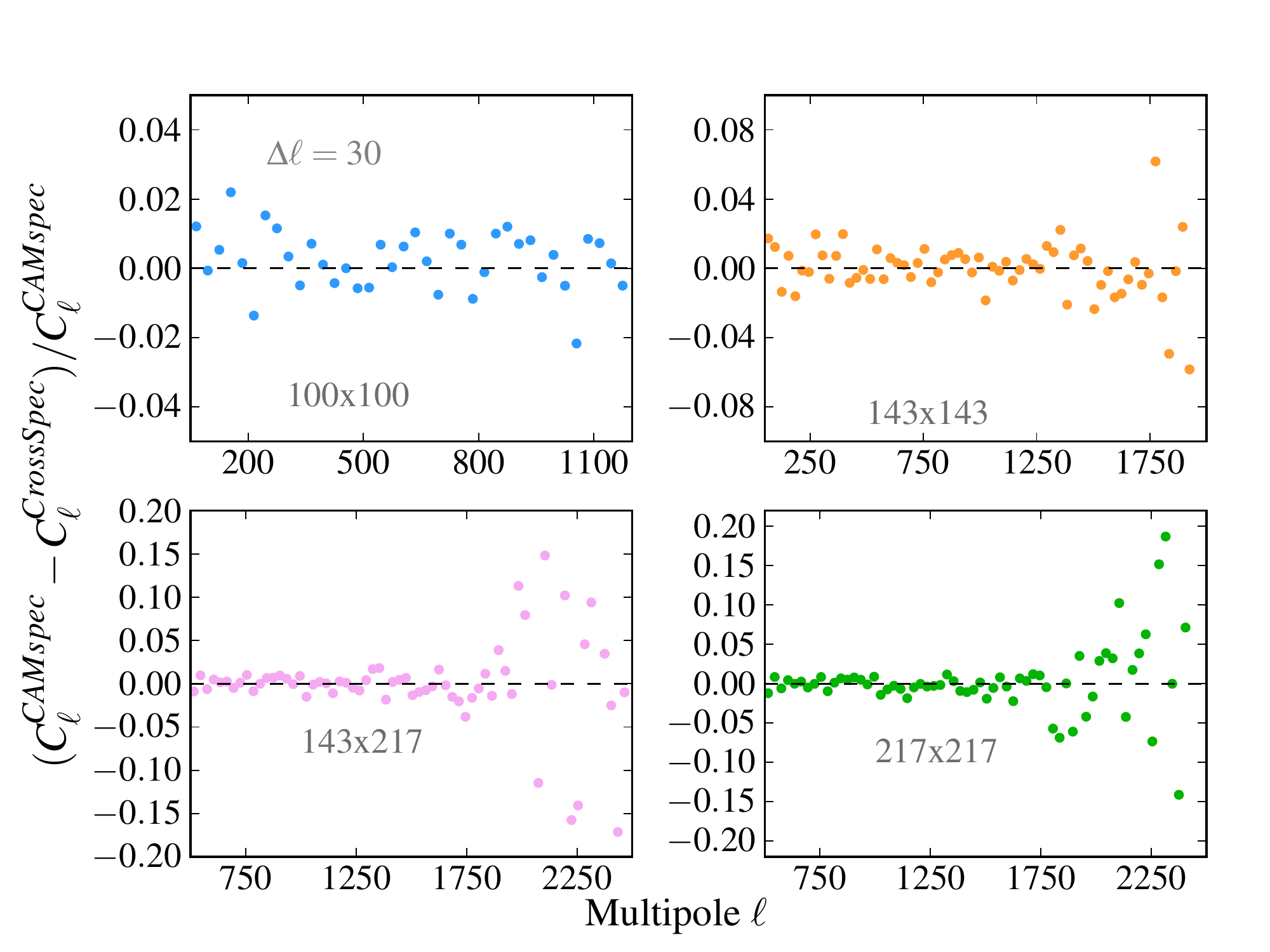}\\ [0.0cm]
\end{array}$
    \caption{Season cross and detector set spectra: the Planck CAMspec spectra compared to those computed from the season maps, but using the same masking and galactic foreground modelling procedure as outlined in the Planck likelihood paper. The $\ell\sim1800$ feature is not present in the cross season spectra even without any additional cleaning procedure applied. \label{fig:specCompare}}
  \end{center}
\end{figure*}

The spectra used in the Planck high-$\ell$ likelihood were made using data from individual detector sets. While these spectra have reduced scatter from the many cross correlations possible between detector sets, they are also more likely to be prone to any systematic effect correlated between detectors, such as the 4K cooler line and the notch filter that produces echoes in the map (see Section  3.6 in~\citep{planck:hfi}). 
\begin{table*} [ht]
\caption{Table of mean values of parameters from the nominal Planck data. \label{table:analysis}}
\begin{center}
\begin{tabular}{|c|c|c|c|c|}
\hline
                    &              &  Uncleaned  &   Half-Ring     & Half-Ring     \\
                    &{Planck Team} & Cross-Season & Survey $1 \cap 2$ & Survey $1 \cup 2$\\
\hline
\hline
$\Omega_c h^2$      & 0.1199        &0.1162        &0.1173          & 0.1186 \\
$n_s$               &0.9603        &0.9698        &0.9655          &0.9595  \\
$h$                 &0.673         &0.688         &0.682           &0.677   \\
100 $\Omega_b h^2$  &2.204         &2.214         &2.197           &2.202   \\
$\log(10^{10} A_s)$ & 3.088        &3.084         &3.087           &3.084   \\
$\tau $             &0.089         &0.091         &0.092           &0.089   \\
 \hline
 $-2\ln\mathcal{L}_\text{CAMspec} $ &7795.11&7462.72&7715.93 &7590.70  \\
 $-2\ln\mathcal{L}_\text{Commander} $ &-6.96&-7.68  &-7.69   &-6.63    \\
 $-2\ln\mathcal{L}_\text{lowlike} $ &2014.44&2014.41&2014.51 &2014.41  \\
 \hline
\end{tabular}
\end{center}
\end{table*}%
In contrast, the publicly available Planck maps are made by combining the data of all detectors in a given channel in the first two Planck observational seasons. The cross-spectra computed from these season maps have larger noise than the detector set cross-spectra but are less sensitive to such systematics. A comparison between the detector set spectra and the season cross-spectra is shown in Figure~\ref{fig:specCompare}. 

As shown in table~\ref{table:analysis}, the cosmological parameters derived from the detector set spectra are systematically different from the survey cross-spectra using Planck's treatment of foregrounds. The parameters derived from the survey cross-spectra are closer to pre-Planck values of cosmological parameters and are close to the parameters found in our main analysis. Since the survey cross-spectra contain less data than the detector set cross-spectra, some shifts are expected and one cannot immediately conclude that this is due to a systematic. However, as shown in Figure~\ref{fig:nsom}, the cosmological parameters from uncleaned survey cross-spectra agree well with with those derived from cleaned survey cross-spectra for comparable $f_\text{sky}$. For the uncleaned survey cross-spectra we use the same galactic and point-source masks used by Planck. After masking the survey strip, we denote the mask used for the 100 GHz data SA39 and that used for the 143 and 217 GHz data SA24. The figure also shows that parameters derived from cleaned survey cross-spectra show no strong dependence on $f_\text{sky}$. This suggests that these shifts are not purely statistical even in the absence of detailed simulations. 

We have also analyzed the publicly available halfring maps and compare the results of our analysis based on Planck's foreground model with detector set and season cross-spectra in table~\ref{table:analysis}. The halfring cross-spectra contain more data than the survey cross-spectra. However, they contain a small amount of correlated noise that is hard to model. So we do not use them for our main analysis and only use them to check the dependence of cosmological parameters on pixels that were only observed in one of the surveys. As shown in table~\ref{table:analysis}, there is a surprisingly strong sensitivity of cosmological parameters to these pixels. The analysis in which only pixels observed in both surveys are used (Survey $1\cap 2$) leads to parameters close to those found in our main analysis. As pixels only observed in one of the surveys are added (Survey $1\cup2$), the values of the cosmological parameters shift close to those reported by the Planck team.  These pixels would be especially problematic if there were a systematic in one of the surveys. Refining further and including pixels only observed in Survey 1 or pixels only observed in Survey 2, we find that the shift entirely derives from the pixels only observed in Survey 1. In contrast, including pixels only observed in Survey 2 only has a small effect on cosmological parameters. The shifts are too large to be caused by the amount of data added, suggesting either a systematic in these pixels or that the sky happens to be peculiar in the strip only observed in Survey 1.   
\section{Conclusion}

Because of their exquisite sensitivity, the Planck data are our most sensitive measurement of cosmological parameters and
the basic properties of our universe. Because of this sensitivity, it is important that there are independent analyses of the Planck data to complement the work done by the Planck team. 

We have performed such an independent analysis and have found that the $217\times217$ data drives the tension between cosmological parameters determined from the CMB and those determined from astronomical measurements.  Re-analyzing the Planck maps and using only the season cross-spectra, we find cosmological parameters more consistent with other measurements.  This suggests that the tension is at least in part due to cross-correlations between 217 detectors observing the sky at the same time.  Whether this is a statistical fluke or a signature of a residual systematic in the 217 GHz maps is difficult to determine with the current data.  However, the upcoming release and analysis of the full five seasons of Planck data should clarify this issue and provide even more detailed insights into the basic properties of our universe.

\acknowledgements{We thank members of the Planck team for their help and advice on using the remarkable public data and for their efforts in making the data derived from their many years of work available to the broader cosmology community. They have offered much advice and support in understanding the various data products. We would like to thank Marco Bersanelli, Erminia Calabrese, Olivier Dor{\'e}, Aur{\'e}lien Fraisse, Duncan Hanson, Antony Lewis, Marius Millea, Hiranya Peiris, Jean-Loup Puget, Paul Shellard, and in particular Anthony Challinor, George Efstathiou, Dick Bond and Jo Dunkley. We gratefully acknowledge the use of the HEALpix package~\citep{Gorski:2004by}. DNS' work on this project was supported partially by NSF and NASA. RF would like to thank the Raymond and Beverly Sackler Foundation for their support. RF is also supported in part by the NSF under grants NSF-PHY-1213563 and NSF-PHY-0645435.}

\bibliographystyle{apsrev1}
\bibliography{planck}

\end{document}